# A rigorous coupled-wave analysis of birefringent holographic gratings with periodically-modulated dielectric tensor along an in-plane direction and tensor variations in the thickness direction

Masud Mansuripur[*]† and Per Kristen Jakobsen‡

†James C. Wyant College of Optical Sciences, The University of Arizona, Tucson

‡Department of Mathematics and Statistics, The Arctic University of Norway, Tromsø, Norway

**Abstract**. Diffraction of light upon interaction with thick slabs of a dielectric material having a periodic modulation of its refractive index (or dielectric tensor) is typically studied with the aid of the method known as the rigorous coupled-wave analysis (RCWA). The method involves solving Maxwell's equations for a large number of coupled electromagnetic plane-waves inside the dielectric slab, then matching the boundary conditions at the interface between the incidence medium and the slab, as well as those at the interface between the slab and the transmittance medium. In this way, one obtains the $E$-field and $H$-field amplitudes for all the reflected and transmitted plane-waves (i.e., diffraction orders as well as evanescent waves) that emerge within the incidence and transmittance media. If the refractive index (or dielectric tensor) of the holographic slab happens to vary in the thickness direction, one treats the slab as a number of thin layers stacked upon each other, then computes and combines the scattering matrices of these layers to arrive at the complete solution for the entire stack. The goal of the present paper is to extend the standard RCWA method to the case where the hologram's dielectric tensor varies in the thickness direction (in addition to being periodically modulated along an in-plane axis), *without* slicing up the thick hologram into a number of thin layers. The reflected and transmitted plane-waves in this case exhibit a large degree of degeneracy, but our numerical results confirm the validity and the accuracy of our proposed algorithm for handling such degeneracies.

**1. Introduction**. The rigorous coupled-wave analysis (RCWA) of diffraction from thick holographic slabs has a long and distinguished history.[1-18] In essence, the analysis involves solving Maxwell's equations for a large (but finite) number of plane-waves inside a holographic slab, whose refractive index (or dielectric tensor) is periodically-modulated in the $xy$-plane of the slab. The periodicity of the optical characteristics of the holographic medium allows one to represent the slab via a Fourier series, the individual components of which are then responsible for the coupling between pairs of plane-waves that reside within the slab. The linear equations that relate the properties of the various plane-waves to each other (via Maxwell's equations and also the Fourier series coefficients of the hologram) are arranged in the form of a large matrix equation, whose eigenvalues and eigenvectors represent the essential features of the complete set of coupled plane-waves. Modern computers are quite efficient at finding these eigenvalues and eigenvectors numerically, thus revealing the fundamental characteristics of the individual plane-waves that, collectively, satisfy Maxwell's equations in the presence of the periodically-modulated dielectric tensor of the holographic slab.

The aforementioned eigenvectors of the large matrix need proper scaling, as each remains indeterminate to within a multiplicative factor. The heretofore unspecified coefficients of these eigenvectors are subsequently fixed by matching the boundary conditions at the entrance and exit facets of the slab. Aside from an incident plane-wave that propagates toward the slab within the medium of incidence, there exist numerous other plane-waves (of both propagating and evanescent types) within the incidence and transmittance media that also need to be specified. Enforcing the continuity of the tangential components of the electric and magnetic fields ($\boldsymbol{E}_\parallel$ and $\boldsymbol{H}_\parallel$) at the slab boundaries provides precisely as many linear equations as there are unknown parameters in this system of equations. Thus, upon matching the boundary conditions at the entrance and exit facets of the slab, one obtains a large matrix equation for the set of unknown coefficients that can, once again, be solved numerically, this time by inverting a large square matrix. In this way, all the relevant equations of Maxwell are solved consistently, and the reflected and transmitted plane-waves (commonly referred to as diffraction orders), as well as the entire set of coupled plane-waves inside the slab, are fully identified.

[*]Corresponding author's email address: masud@optics.arizona.edu.

It is not our aim here to review the electromagnetic (EM) theory behind the RCWA, nor its various numerical implementations or experimental verifications. Our specific goal is to propose a method (culminating in an algorithm) for the computation of the diffraction efficiencies of various reflected and transmitted orders emerging from a holographic slab whose dielectric tensor $\tilde{\varepsilon}$ has a periodic variation (i.e., modulation) in the $xy$-plane of the slab, say, along the $x$-axis, in addition to having some form of variation across the thickness (i.e., along the $z$-axis). Traditionally, this type of problem has been solved by slicing the thick hologram into a number of thin layers, computing the scattering matrices of the individual layers, then combining these matrices to arrive at the scattering matrix of the thick hologram in its entirety. The proposed method of the present paper does away with the slicing and, instead, incorporates the complete set of two-dimensional (i.e., $xz$) Fourier series coefficients of the slab's dielectric tensor into a rigorous coupled-wave analysis of the holographic grating. Our procedure introduces, by necessity, a large degeneracy into the set of reflected and transmitted diffraction orders, which we then accommodate by treating each diffracted order as a coherent superposition of many co-propagating (or spatially coinciding) orders.

Our non-magnetic hologram has a base refractive index $n_2 = \sqrt{\varepsilon_2}$ that is augmented and modulated by a dielectric tensor $\tilde{\varepsilon}_h(x, z)$, whose natural axes of biaxial birefringence are allowed to have an arbitrary orientation within the $xyz$ space occupied by the holographic slab. The slab is confined to the region between its upper facet in the $z = 0$ plane, and its lower facet in the $z = -d$ plane. The homogeneous, isotropic, transparent, linear, and semi-infinite media of incidence and transmittance have real-valued and positive refractive indices $n_1 = \sqrt{\varepsilon_1}$ and $n_3 = \sqrt{\varepsilon_3}$, respectively. The incident plane-wave is monochromatic, having frequency $\omega$ and free-space wavelength $\lambda_0 = 2\pi c/\omega$ (where $c$ is the speed of light in vacuum); it arrives at the upper facet of the hologram at oblique incidence, specified by its $k$-vector's polar and azimuthal angles $(\theta, \varphi)_{\text{inc}}$. The reflected diffracted orders return to the incidence medium, while the transmitted orders emerge from the exit facet of the holographic slab and propagate within the transmittance medium. In addition to the propagating reflected and transmitted beams (i.e., plane-waves), there reside many evanescent plane-waves immediately above and also immediately below the holographic slab. These evanescent waves, of course, do not carry any EM energy away from the slab; nevertheless, their existence is essential for matching the boundary conditions (i.e., ensuring the continuity of the tangential $\boldsymbol{E}$ and $\boldsymbol{H}$ fields) at the slab's upper and lower facets.

The organization of the paper is as follows. In Sec.2, we provide a broad overview of the diffraction problem under consideration, where we describe the general setup of our optical system. The periodic modulation of the dielectric tensor in the $xy$-plane of the hologram, in combination with the tensor's variation in the thickness direction (i.e., along the $z$-axis), allows for numerous interesting configurations, some of which are briefly discussed in the examples at the end of the section. A detailed account is given in Sec.3 of the reflection and transmission efficiencies of the various diffracted orders obtained by a numerical implementation of our RCWA for rectangular and trapezoidal grating profiles. The computed results (involving thousands of coupled plane-waves inside as well as outside the slab under oblique illumination) confirm the accuracy and the stability of the algorithm in the presence of optical birefringence within the hologram. In the absence of optical absorption across the material system, a crucial test of the validity of our algorithm is the conservation of the EM energy among the incident, reflected, and transmitted beams. The reported results in Sec.3, a small sampling of the numerous cases that we have examined, confirm the stability of the algorithm and the conservation of optical energy in each and every instance.

Section 4 is devoted to presenting the coupled-wave theory of our two-dimensional diffraction gratings — i.e., those with a periodic modulation of the dielectric tensor $\tilde{\varepsilon}$ in the $xy$-plane of the slab along the $x$-axis, in addition to an arbitrary variation of the tensor in the thickness direction along the $z$-axis. This section also includes a step-by-step recipe for the numerical implementation of the proposed algorithm. The paper closes in Sec.5 with a few concluding remarks and some general observations.

Certain theoretical details are relegated to four appendices at the end of the paper. Appendix A describes the hologram's dielectric tensor when its natural (generally biaxial) birefringence axes are rotated through the Euler angles $(\theta,\varphi,\chi)_h$ in the three-dimensional $xyz$ space of the holographic medium. Appendix B provides a detailed account of the two-dimensional Fourier transformation of a general-purpose trapezoidal grating profile. In Appendix C, we elaborate the matching of the boundary conditions at the top facet of the hologram (i.e., the $z = 0$ plane), and also that at the hologram's bottom surface (i.e., the $z = -d$ plane). Appendix D is devoted to a calculation of the Poynting vector for the various plane-waves that reside in the incidence and transmittance media; this pertains to the computation of the EM energy flux for the incident plane-wave as well as that of the reflected and transmitted plane-waves that constitute the various diffraction orders.

**2. Overview and general set-up of the problem**. With reference to Fig.1, we define the dielectric tensor profile inside a holographic slab of thickness $d$ as $\tilde{\varepsilon}_2(x,z) = \varepsilon_2\tilde{I} + g(x,z)\tilde{\varepsilon}_h$. The function $g(x,z)$ is periodic along the $x$-axis, with a period of $\Lambda$. The dielectric tensor $\tilde{\varepsilon}_h$ is rotated within the $xyz$ coordinate system through the Euler angles $(\theta,\varphi,\chi)_h$, which are defined in Appendix A. The incidence and transmittance media are isotropic, homogeneous, linear, and transparent, having dielectric constants $\varepsilon_1$ and $\varepsilon_3$, respectively — these correspond to real-valued positive refractive indices $n_1 = \sqrt{\varepsilon_1}$ and $n_3 = \sqrt{\varepsilon_3}$. The monochromatic incident plane-wave has frequency $\omega$, free-space wavelength $\lambda_0 = 2\pi c/\omega$, polar and azimuthal incidence angles $(\theta,\varphi)_{\text{inc}}$, and polarization state $(E_p,E_s)_{\text{inc}}$, specified via the complex-valued amplitudes of the $p$ and $s$ components of the incident $E$-field. (By definition, $\boldsymbol{E}_p$ and $\boldsymbol{E}_s$ are in the plane and perpendicular to the plane of incidence, respectively, with the plane of incidence containing the incident $k$-vector and the $z$-axis. Note that the true polar angle of incidence $\theta$ within our $xyz$ coordinate system is the supplementary angle to that depicted in Fig.1.)

We proceed to define $K_x = 2\pi/\Lambda$ and $K_z = 2\pi/d$, then expand the function $g(x,z)$ in a truncated two-dimensional Fourier series containing a total of $(2M+1)(2N+1)$ terms, as follows:

$$g(x,z) = \textstyle\sum_{m=-M}^{M}\sum_{n=-N}^{N} g_{mn}\exp[\mathrm{i}(mK_x x + nK_z z)], \tag{1a}$$

$$g_{mn} = (\Lambda d)^{-1}\int_{z=-d}^{0}\int_{x=0}^{\Lambda} g(x,z)\exp[-\mathrm{i}(mK_x x + nK_z z)]\,\mathrm{d}x\mathrm{d}z. \tag{1b}$$

If the function $g(x,z)$ happens to be expressible as $f(ax+bz)$, a periodic function with period $L$ along the vector $\boldsymbol{c} = a\hat{\boldsymbol{x}} + b\hat{\boldsymbol{z}}$, there will be no need to treat the holographic slab as a two-dimensional grating, since the standard RCWA method for one-dimensional gratings can be readily employed in this case. To see this, write $\boldsymbol{c} = \sqrt{a^2+b^2}(\sin\vartheta\,\hat{\boldsymbol{x}} + \cos\vartheta\,\hat{\boldsymbol{z}})$, then observe that $\vartheta$ is the angle between $\boldsymbol{c}$ and the $z$-axis. The grating vector will then be $\boldsymbol{K} = (2\pi/L)(\sin\vartheta\,\hat{\boldsymbol{x}} + \cos\vartheta\,\hat{\boldsymbol{z}})$ as depicted in Fig. 2. The index-modulation function may now be expressed in terms of $\boldsymbol{K} = K_x\hat{\boldsymbol{x}} + K_z\hat{\boldsymbol{z}}$, as follows:

$$g(x,z) = \sum_{n=-\infty}^{\infty} g_n e^{\mathrm{i}n(K_x x + K_z z)}, \qquad \left[g_n = L^{-1}\int_0^L f(\zeta)e^{-\mathrm{i}2\pi\zeta/L}\mathrm{d}\zeta\right]. \tag{2}$$

It should be clear that the grating period $\Lambda$ in the $x$-direction is related to the period $L$ along the $K$-vector via $\Lambda = L\sin\vartheta$.

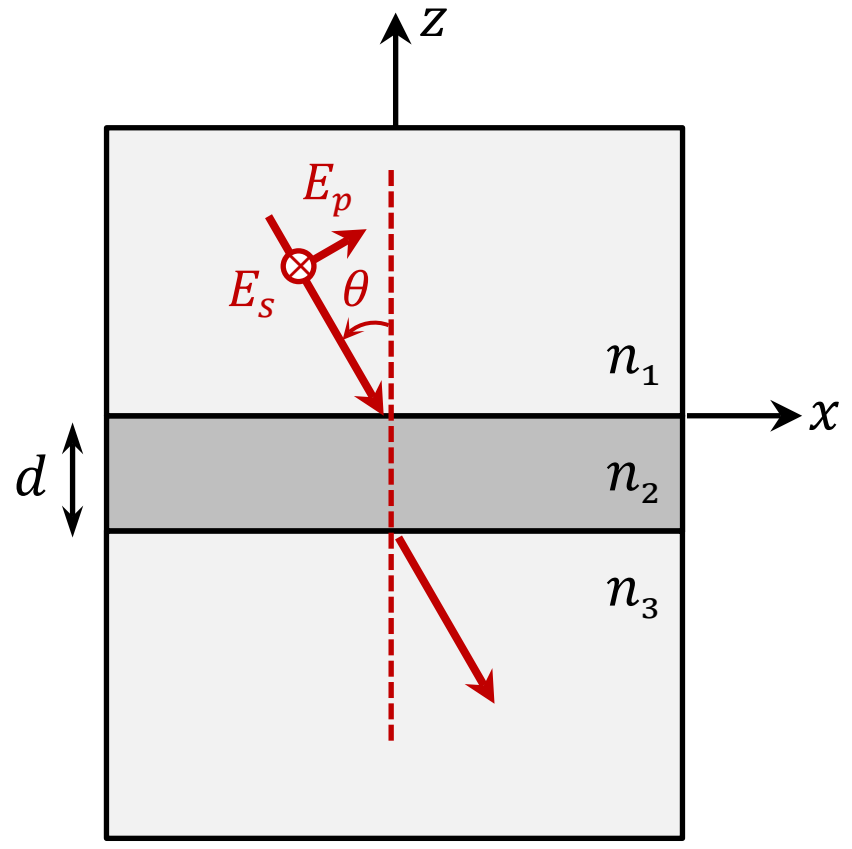

**Fig. 1**. A monochromatic plane-wave of frequency $\omega$ arrives at oblique incidence at the interface between a transparent medium of refractive index $n_1$ and a slab of thickness $d$, whose base refractive index is $n_2$. The transmission medium on the backside of the slab is also transparent, with a refractive index of $n_3$. The polar and azimuthal angles of incidence are $(\theta,\varphi)_{\text{inc}}$, with the incident beam having an $E$-field component $\boldsymbol{E}_p$ in the plane of incidence, and another component $\boldsymbol{E}_s$ perpendicular to that plane. (**Note**: The true polar angle of incidence $\theta$ is the supplementary angle to that depicted in the figure.) The slab's base refractive index $n_2$ is augmented by a dielectric tensor $\tilde{\varepsilon}_h$, which can be an arbitrary function of $x$ and $z$, albeit with periodic repeats (period $= \Lambda$) along the $x$-axis; that is, $\tilde{\varepsilon}_h(x,z) = \tilde{\varepsilon}_h(x+\Lambda,z)$.

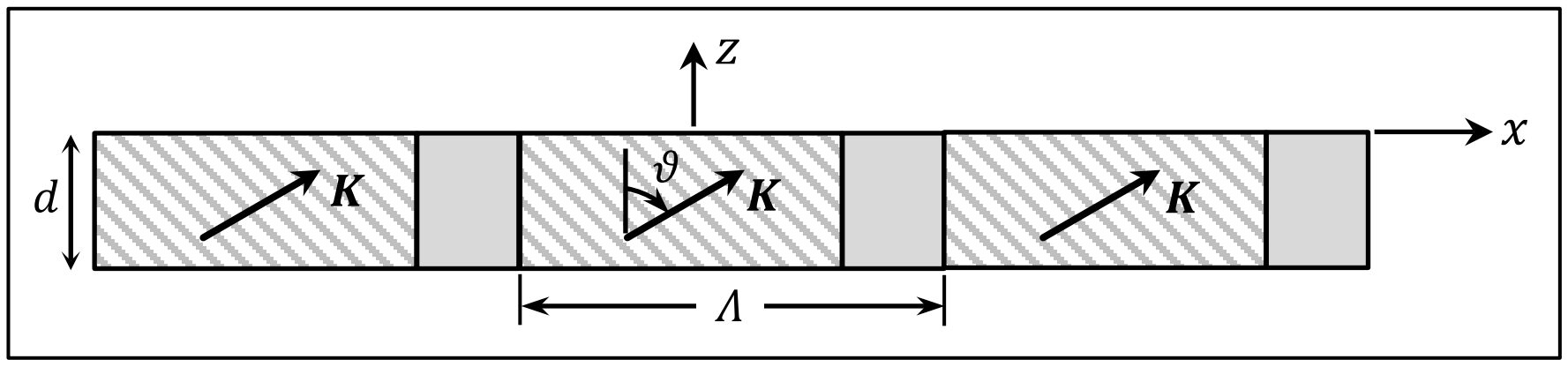

**Fig. 2**. If the holographic slab's index-modulation is expressible in terms of a function of a single variable in the form of $g(x,z) = f(ax+bz)$, then the vector $\boldsymbol{c} = a\hat{\boldsymbol{x}} + b\hat{\boldsymbol{z}} = \sqrt{a^2+b^2}(\sin\vartheta\,\hat{\boldsymbol{x}} + \cos\vartheta\,\hat{\boldsymbol{z}})$ would define the direction along which the dielectric tensor varies periodically. Denoting by $L$ the period of the index modulation along $\boldsymbol{c}$, one can define a grating-vector $\boldsymbol{K} = (2\pi/L)(\sin\vartheta\,\hat{\boldsymbol{x}} + \cos\vartheta\,\hat{\boldsymbol{z}})$. Considering that $\Lambda$ is the period of the holographic grating along the $x$-axis, we will have $L = \Lambda/\sin\vartheta$.

---

**Example 1**. As an example of a genuinely two-dimensional holographic grating, consider the slab depicted in Fig. 3(a), whose index modulation is confined to a rectangular region of length $\alpha\Lambda$ and width $\beta d$, with $\alpha$ and $\beta$ being in the $(0,1)$ interval. The function $g(x,z)$ and its Fourier coefficients in this case are given by

$$g(x,z) = \begin{cases} 1, & \text{when } ½\Lambda(1-\alpha) \le x \le ½\Lambda(1+\alpha) \text{ and } -½d(\beta+1) \le z \le ½d(\beta-1); \\ 0, & \text{otherwise.} \end{cases} \tag{3}$$

$$g_{mn} = (\Lambda d)^{-1}\int_{z=-½d(\beta+1)}^{½d(\beta-1)}\int_{x=½\Lambda(1-\alpha)}^{½\Lambda(1+\alpha)} e^{-\mathrm{i}(mK_x x + nK_z z)}\mathrm{d}x\mathrm{d}z = \left[\frac{\sin(\pi m\alpha)}{\pi m}\right]\left[\frac{\sin(\pi n\beta)}{\pi n}\right]e^{\mathrm{i}\pi(n-m)}. \tag{4}$$

As an alternative index profile, let $g(x, y)$ be 1.0 inside, and 0.0 outside, a circle of radius $R$, with $2R \le \min(\Lambda, d)$, as shown in Fig. 3(b). In this case, the Fourier coefficients $g_{mn}$ will be

$$g_{mn} = (\Lambda d)^{-1} \iint_{\text{circle}} e^{-\mathrm{i}(mK_x x + nK_z z)} \mathrm{d}x\mathrm{d}z$$
$$= R J_1\left(2\pi\sqrt{(mR/\Lambda)^2 + (nR/d)^2}\right) e^{\mathrm{i}\pi(n-m)} / \sqrt{m^2 d^2 + n^2 \Lambda^2}. \tag{5}$$

Here, $J_1(\cdot)$ is a Bessel function of the first kind, order 1.

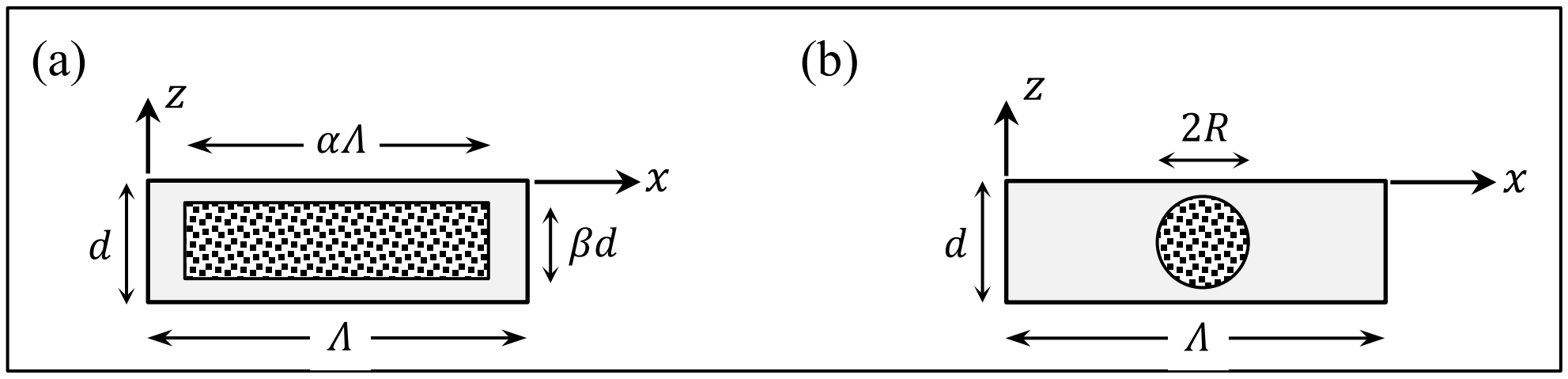

**Fig. 3**. A single period of a holographic grating, having thickness $d$ in the $z$-direction and period $\Lambda$ along the $x$-axis. The base refractive index $n_2$ of the holographic medium is augmented by a dielectric tensor $\tilde{\varepsilon}_h$ within the shaded area at the center of each period of the slab. (a) The length and width of the index-modulated rectangular region are $\alpha\Lambda$ and $\beta d$, with $\alpha$ and $\beta$ being in the $(0, 1)$ interval. (b) The radius of the index-modulated circular region at the center of each period of the slab is $R$.

---

**Example 2**. As another elementary example, consider the holographic grating shown in Fig. 4, whose dielectric tensor is modulated sinusoidally in the thickness direction, albeit with different periods in different segments of the $x$-axis. Let the functional form of $g(x, z)$ be

$$g(x, z) = \sin(2\pi n_1 z/d) \text{ when } 0 \le x < \alpha\Lambda \text{ and } \sin(2\pi n_2 z/d) \text{ when } \alpha\Lambda \le x < \Lambda. \tag{6}$$

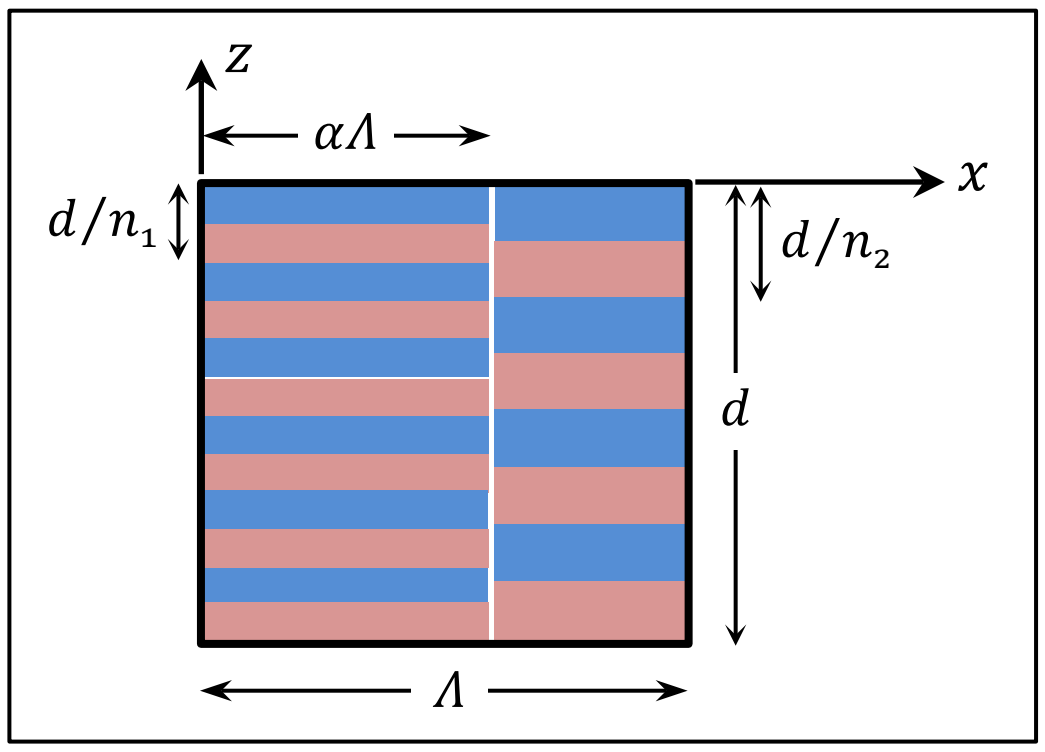

**Fig. 4**. The function $g(x, z)$ of Eq.(6) corresponds to a holographic grating with thickness $d$ in the $z$-direction and period $\Lambda$ along the $x$-axis. The holographic slab's dielectric tensor is sinusoidally modulated in the $z$-direction, albeit with different periodicities in different segments of the $x$-axis. Whereas for $0 \le x < \alpha\Lambda$ the modulation period is $d/n_1$, that for $\alpha\Lambda \le x < \Lambda$ is $d/n_2$.

The Fourier coefficients of $g(x, z)$ will then be

$$g_{mn} = (2\mathrm{i}\Lambda d)^{-1}\left\{\int_{x=0}^{\alpha\Lambda} e^{-\mathrm{i}2\pi mx/\Lambda}\mathrm{d}x \int_{z=-d}^{0}\left[e^{\mathrm{i}2\pi(n_1-n)z/d} - e^{-\mathrm{i}2\pi(n_1+n)z/d}\right]\mathrm{d}z\right.$$
$$\left. + \int_{x=\alpha\Lambda}^{\Lambda} e^{-\mathrm{i}2\pi mx/\Lambda}\mathrm{d}x \int_{z=-d}^{0}\left[e^{\mathrm{i}2\pi(n_2-n)z/d} - e^{-\mathrm{i}2\pi(n_2+n)z/d}\right]\mathrm{d}z\right\}$$
$$= \mathrm{i}e^{-\mathrm{i}m\pi\alpha}\frac{\sin(m\pi\alpha)}{2m\pi}(\delta_{n,n_2} - \delta_{n,n_1} - \delta_{n,-n_2} + \delta_{n,-n_1}). \tag{7}$$

Here, $\delta_{n,\pm n_1}$ and $\delta_{n,\pm n_2}$ are Kronecker deltas.

---

**Example 3**. Figure 5 shows the cross-section of a holographic grating with a trapezoidal profile for the function $g(x, z)$, which equals 1.0 in the dotted regions and 0.0 in the uniformly grey regions of the slab. Each period of the grating is a long strip along the $y$-axis, having width $\Lambda$ along $x$ and thickness $d$ along the $z$-axis. In the grey regions, the refractive index is $n_2 = \sqrt{\varepsilon_2}$, whereas in the dotted regions the base refractive index is augmented by a dielectric tensor $\tilde{\varepsilon}_h$, so that the overall tensor within each dotted trapezoid is $\varepsilon_2 \tilde{I} + \tilde{\varepsilon}_h$. Appendix B provides a detailed derivation of the Fourier coefficients $g_{mn}$ of a trapezoidal function $g(x, z)$.

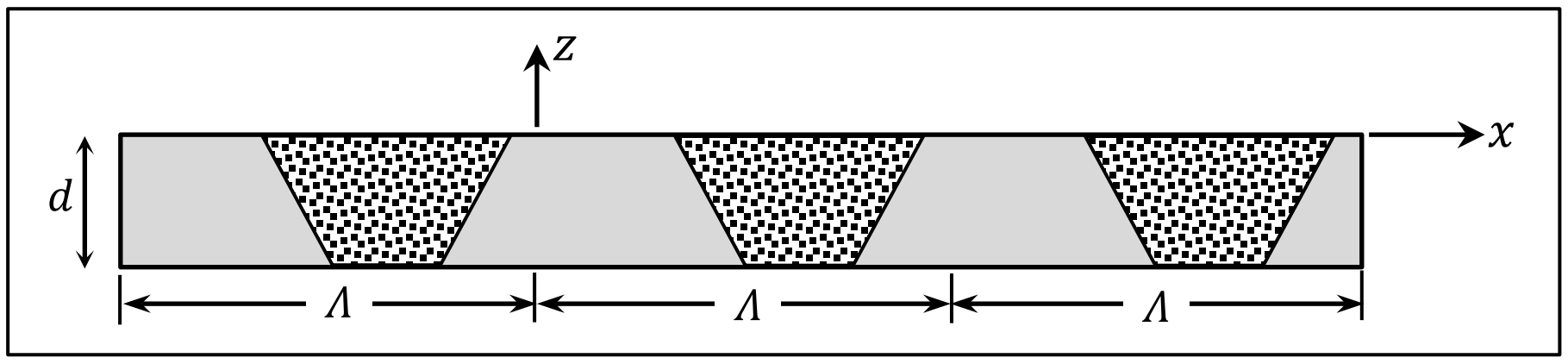

**Fig. 5**. Cross-sectional profile of a holographic grating having thickness $d$ along $z$ and period $\Lambda$ along the $x$-axis. Within the dotted trapezoidal regions, the base refractive index $n_2 = \sqrt{\varepsilon_2}$ of the slab material is augmented by a dielectric tensor $\tilde{\varepsilon}_h$. The overall dielectric tensor of the slab is $\varepsilon_2 \tilde{I} + \tilde{\varepsilon}_h\, g(x, z)$, with the function $g(x, z)$ being 1.0 inside the dotted trapezoids, and 0.0 in the uniformly grey regions.

If necessary, one may choose to introduce artificial features into the function $g(x, z)$ *outside* its proper range of $-d \le z \le 0$, in order to eliminate sharp discontinuities (along $z$) that would otherwise give rise to the well-known Gibbs phenomenon[19] of the Fourier transform theory. Introducing smooth transitions-to-zero in the regions $-d_1 \le z \le -d$ and $0 \le z \le d_2$, as shown in Fig.6, will *not* alter the desired $g(x, z)$ inside its proper range of $-d \le z \le 0$, but it *does* change the "effective" grating vector along $z$ to $K_z = 2\pi/(d_1 + d_2)$. Smoothing out the transitions-to-zero of $g(x, z)$ in the regions above and below the holographic slab, and ensuring that $d_1 + d_2$ is only slightly greater than the thickness $d$, will also help to reduce the overall number of the Fourier coefficients $g_{mn}$ that are needed to accurately represent the function $g(x, z)$.

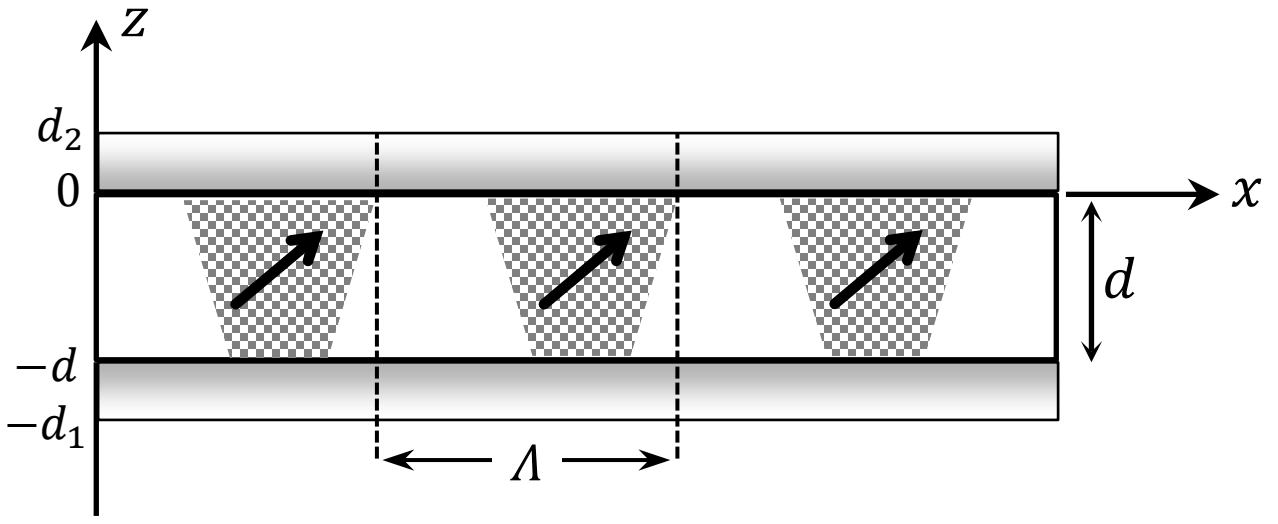

**Fig.6**. The periodically repeating function $g(x, z)$ equals 1.0 in areas where the holographic slab is birefringent, and zero outside the birefringent regions, where the hologram is homogeneous. Sharp discontinuities along the $z$-axis that might arise from differences between $g(x, 0)$ and $g(x, -d)$ can be smoothed out by introducing artificial transition regions above and below the slab, i.e., in the regions $-d_1 \le z < -d$ and $0 < z \le d_2$. Within these transition regions, the function $g(x, z)$ declines smoothly (but otherwise arbitrarily) from its upper and lower boundary values (i.e., values at $z = 0$ and $z = -d$), eventually vanishing (again smoothly) as $z$ approaches $-d_1$ and $d_2$, respectively.

---

In general, any profile of the refractive index (or dielectric tensor) within a strip of thickness $d$ and width $\Lambda$ that is periodically repeated along the $x$-axis is admissible for a holographic slab. The Fourier coefficients of the index profile can be computed analytically or numerically, as the case may be. One then uses a finite number, say, $(2M + 1)(2N + 1)$, of these Fourier coefficients to represent, to good approximation, the index-modulation of the holographic slab.

The standard techniques of RCWA are subsequently used to obtain the diffraction efficiencies of the various reflected and transmitted orders that emerge within the incidence and transmittance media. The only difficulty in applying the standard techniques arises when one attempts to match the boundary conditions at the interface between media 1 and 2, and also those between media 2 and 3. The reason for this difficulty is that many more (coupled) plane-waves now exist within the holographic slab (medium 2) than are available within media 1 and 3 for matching the boundary conditions at the upper and lower interfaces. The trick is to allow each diffracted order — be it a reflected plane-wave residing in medium 1, or a transmitted plane-wave in medium 3 — to have a degree of degeneracy commensurate with the corresponding number of available plane-waves residing in medium 2. In the end, all the degenerate plane-waves representing a single reflected, or a single transmitted, diffracted order must be coherently combined (i.e., their $E$-field amplitudes, and, similarly, their $H$-field amplitudes, must be added together) to arrive at each diffracted order that emerges within medium 1 (i.e., reflected order), or within medium 3 (i.e., transmitted order).

When matching the boundary conditions at the upper interface (i.e., that between the incidence medium and the holographic slab), care must be taken regarding the zeroth-order reflected beam. Here, the zeroth-order beam is the specularly reflected plane-wave that pairs with the incident plane-wave when implementing the boundary conditions at the upper interface. Because of the aforementioned degeneracy of the zeroth-order reflected beam, one must introduce a similar degeneracy into the incident plane-wave in order to generate the requisite number of boundary conditions at the upper interface. This can be done by breaking up the incident plane-wave into several degenerate (i.e., co-propagating) incident plane-waves whose coherent superposition would reproduce the incident wave. The breakup of the incident wave can be done arbitrarily, and the final results should be independent of how one chooses to implement the breakup.

**3. Results and discussion**. We begin by examining a one-dimensional holographic grating whose index-modulation profile is described by a simple rectangular function with no $z$-dependence; that is, $g(x,z) = \text{rect}(2x/\Lambda)$. Here, $\text{rect}(\zeta)$ is a standard rectangular function that equals 1.0 when $|\zeta| \leq$ ½, and 0.0 otherwise.[19] Figure 7(a) shows a plot of $\text{rect}(2x/\Lambda)$ versus $x$ over one period $\Lambda$ of the grating; superimposed on the rectangular "groove" profile is an approximate version of the groove, constructed from its first 49 Fourier coefficients (i.e., $M = 24$). The base refractive index of the holographic slab was set to $n_2 = 1.5$, and the dielectric tensor within the groove was chosen to be a diagonal matrix $(\varepsilon_{h1}, \varepsilon_{h2}, \varepsilon_{h3}) = (0.06, 0.09, 0.02)$, rotated through the Euler angles $(\theta, \varphi, \chi)_h = (45°, 30°, 36°)$. The slab thickness is $d = 100$ µm, the grating period is $\Lambda = 1.5$ µm, the incidence and transmittance media have refractive indices $n_1 = n_3 = 1.0$, and the linearly-polarized incident beam has $(E_p, E_s) = (2/\sqrt{3}, 1)$, free-space wavelength $\lambda_0 = 0.505$ µm, and polar and azimuthal incidence angles $(\theta, \varphi)_{\text{inc}} = (150°, 0°)$.

Figures 7(b) and 7(c) show the computed reflection and transmission diffraction efficiencies for the various diffraction orders. The dominant order is seen to be the $-3^{\text{rd}}$-order transmitted beam, with small fractions of the incident light also thrown into the $0^{\text{th}}$-order transmitted, as well as the $-3^{\text{rd}}$-order and $0^{\text{th}}$-order reflected, beams. The overall diffraction efficiency of the hologram is 89%, with the remaining 11% divided between the $0^{\text{th}}$-order reflected and $0^{\text{th}}$-order transmitted beams. In these calculations, the total number of (coupled) plane-waves inside the holographic slab was $(2M+1)(2N+1) = 31 \times 31 = 961$, which is an overkill, of course, considering that the index-modulation profile is independent of $z$. (The same results were obtained when we set $N = 0$.) The numerical results did not change perceptibly when we increased the value of $M$ beyond this point.

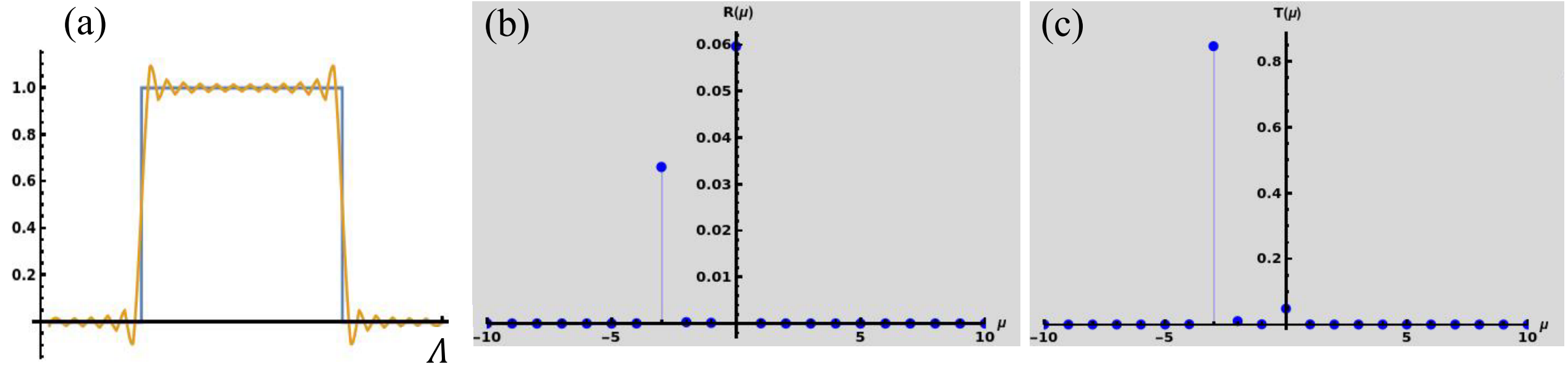

**Fig. 7**. (a) Rectangular groove profile (blue) and its approximate version (orange) computed using the first 49 Fourier coefficients of the rectangular function; the grating period along the $x$-axis is $\Lambda = 1.5$ µm. The overshoot and undershoot seen in the vicinity of the sharp transitions in the computed groove profile (i.e., the Gibbs phenomenon[19]) are unavoidable consequences of truncating the Fourier series. (b) Computed diffraction efficiencies for reflected orders. (c) Computed diffraction efficiencies for transmitted orders.

For the remaining simulations, the general geometry of the system under consideration is depicted in Fig.8, which also shows the trapezoidal profile of the index-modulated section of the holographic slab. The incident plane-wave arrives obliquely at $\theta = 30°$ on the upper grating surface. (The true polar angle of incidence is $\theta = 150°$, which is the clockwise deviation angle of the incident $k$-vector away from the $z$-axis.) The plane of incidence is $xz$, and the polarization state of the beam is linear, with $(E_p, E_s) = (2/\sqrt{3}, 1.0)$. Both the incidence and transmittance media are taken to be free space, with $n_1 = n_3 = 1.0$; the base refractive index of the hologram is $n_2 = \sqrt{\varepsilon_2} = 1.5$. The trapezoidal grating profile has period $\Lambda$ and corners at $(\alpha, \beta, \gamma) = (½, ⅔, ⅚)\Lambda$; inside the trapezoid (red region), the dielectric tensor deviates from the base index, namely, $\tilde{\varepsilon}_2(x,z) = \varepsilon_2 \tilde{I} + \tilde{\varepsilon}_h g(x,z)$. In all the reported simulations, the grating is represented by $41 \times 41 = 1681$ Fourier components.

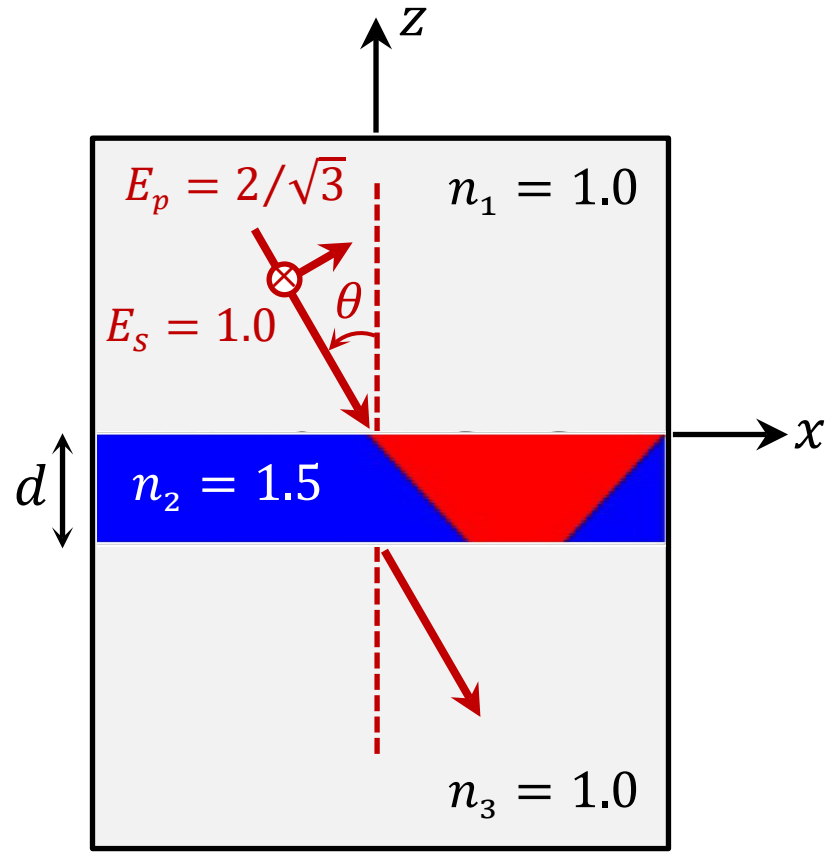

**Fig. 8**. Geometry of the system under consideration. The index-modulated holographic slab has dielectric tensor $\tilde{\varepsilon}_2(x,z) = \varepsilon_2 \tilde{I} + \tilde{\varepsilon}_h g(x,z)$, where the trapezoid function $g(x,z)$ equals 0.0 in the blue region and 1.0 in the red region. In the blue region, the hologram has the base refractive index $n_2 = \sqrt{\varepsilon_2} = 1.5$, while in the red region the overall dielectric tensor is $\varepsilon_2 \tilde{I} + \tilde{\varepsilon}_h$. (**Note**: The true polar angle of incidence, i.e., the clockwise deviation of the incident $k$-vector from the $z$-axis, is the supplementary angle to the $\theta$ depicted here.)

Figure 9 shows the computed reflected and transmitted diffraction efficiencies for a slab whose diagonal dielectric tensor $(\varepsilon_{h1}, \varepsilon_{h2}, \varepsilon_{h3}) = (0.065,\ 0.012,\ 0.021)$ is rotated through the Euler angles $(\theta, \varphi, \chi)_h = (45°, 30°, 36°)$. The slab's thickness $d$ is 100 µm, and the index-modulated trapezoidal profile has period $\Lambda = 15.0$ µm along the $x$-axis. The incident beam has wavelength $\lambda_0 = 0.401$ µm and the incidence angles are $(\theta, \varphi)_{\text{inc}} = (150°, 0°)$. The overall diffraction efficiency of the hologram is 53%, distributed among multiple reflected and transmitted orders, with the remaining 47% divided between the $0^{\text{th}}$-order reflected and $0^{\text{th}}$-order transmitted beams. (**Warning**: The large grating-period-to-wavelength ratio $\Lambda/\lambda_0 \cong 37.4$ in this example allows for 75 propagating diffraction orders in both the incidence and transmittance media. This is substantially greater than the 41 Fourier components used in our simulations. Nevertheless, the diffraction efficiencies reported in Fig.9 seemed to stabilize as we raised the number of Fourier components up to 41 along the $x$-axis.)

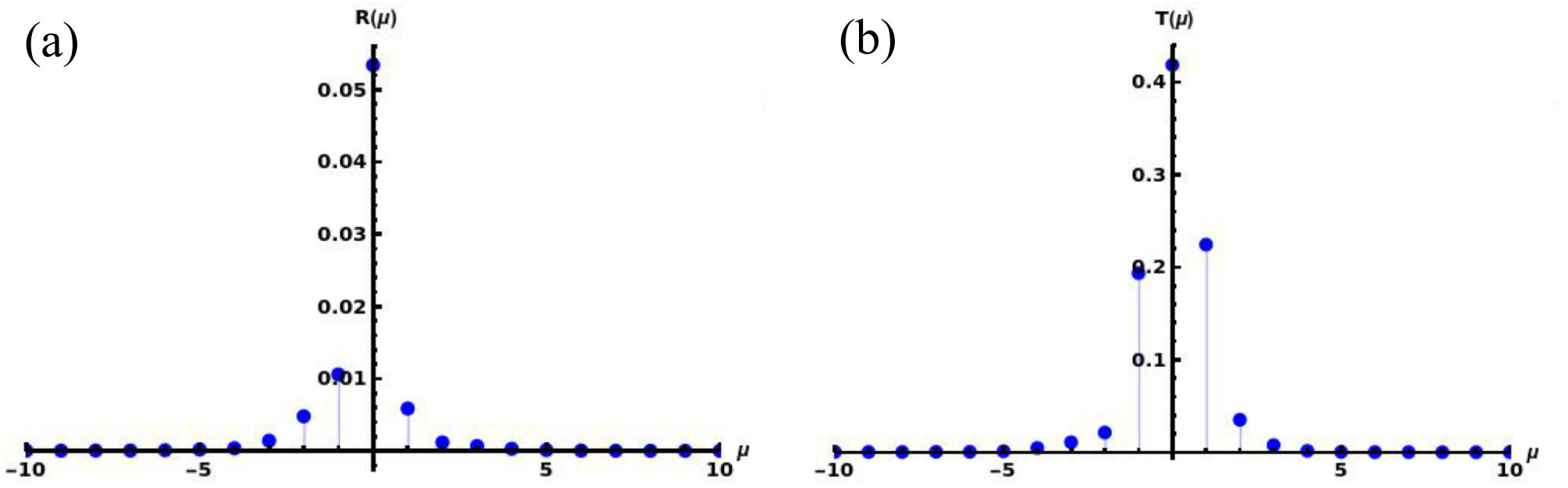

**Fig. 9**. Computed reflected and transmitted diffraction efficiencies for a 100 μm-thick grating with a period of $\Lambda = 15.0$ μm along the $x$-axis. The incident wavelength is $\lambda_0 = 0.401$ μm, and the overall diffraction efficiency is 53%.

Figure 10 shows the computed reflected and transmitted diffraction efficiencies for a slab whose diagonal dielectric tensor $(\varepsilon_{h1}, \varepsilon_{h2}, \varepsilon_{h3}) = (0.065,\ 0.012,\ 0.021)$ is rotated through the Euler angles $(\theta, \varphi, \chi)_h = (45°, 30°, 36°)$. The slab's thickness is $d = 100$ μm, and the period of its index-modulated trapezoidal profile along the $x$-axis is $\Lambda = 1.5$ μm. The incident plane-wave has wavelength $\lambda_0 = 0.506$ μm, and the incidence polar and azimuthal angles are $(\theta, \varphi)_{\text{inc}} = (150°, 0°)$. The dominant diffraction order with 27.5% efficiency is seen to be the $-3^{\text{rd}}$-order transmitted beam. The $0^{\text{th}}$-order reflected beam has 4% efficiency, while 67% of the incident optical power has gone into the $0^{\text{th}}$-order transmitted beam.

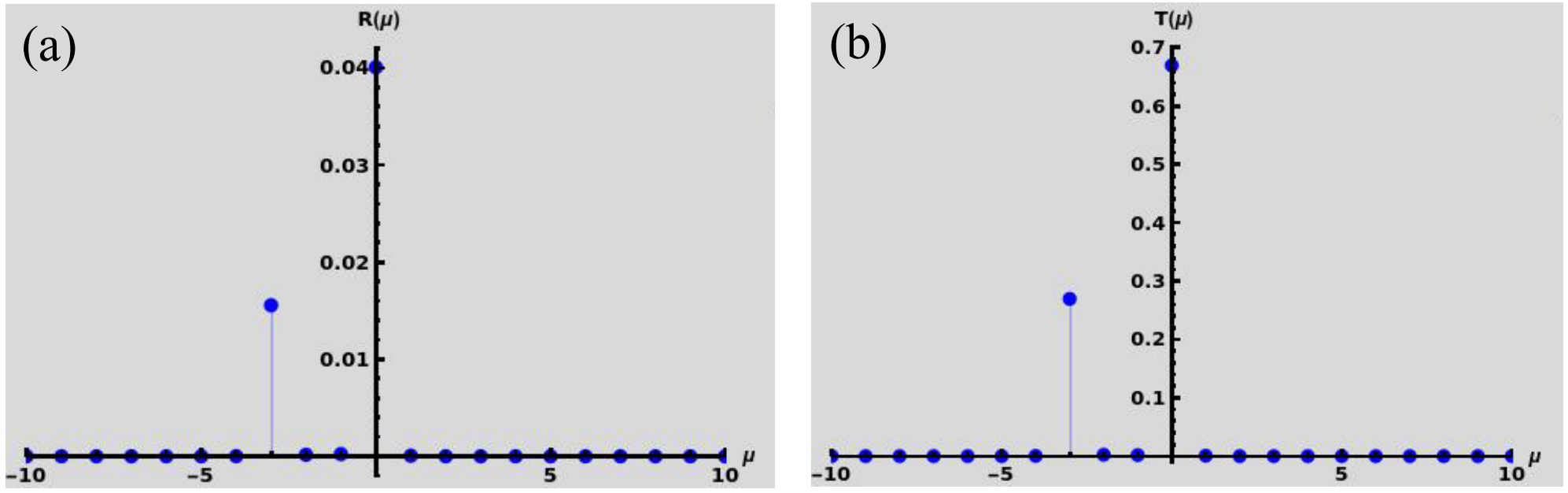

**Fig. 10**. Computed reflected and transmitted diffraction efficiencies for a 100 μm-thick grating with a period of $\Lambda = 1.50$ μm along the $x$-axis. The incident wavelength is $\lambda_0 = 0.506$ μm, and the dominant diffraction orders are the $-3^{\text{rd}}$-order beams with a total efficiency of 29%.

Figure 11 shows the computed reflected and transmitted diffraction efficiencies for a slab whose diagonal dielectric tensor $(\varepsilon_{h1}, \varepsilon_{h2}, \varepsilon_{h3}) = (0.07, 0.1, 0.09)$ is rotated through the Euler angles $(\theta, \varphi, \chi)_h = (45°, 30°, 36°)$. The slab's thickness is $d = 100$ μm, the index-modulated profile is trapezoidal, and the grating period along the $x$-axis is $\Lambda = 0.5$ μm. The incident plane-wave has wavelength $\lambda_0 = 0.497$ μm, and the incidence angles are $(\theta, \varphi)_{\text{inc}} = (150°, 0°)$. The overall diffraction efficiency is 88%, dominated by the $-1^{\text{st}}$-order transmitted beam.

Figure 12 shows the computed reflected and transmitted diffraction efficiencies for a slab whose diagonal dielectric tensor $(\varepsilon_{h1}, \varepsilon_{h2}, \varepsilon_{h3}) = (0.07, 0.02, 0.09)$ is rotated through the Euler angles $(\theta, \varphi, \chi)_h = (45°, 30°, 36°)$. The slab's thickness is $d = 50$ µm, the index-modulation profile is trapezoidal, and the grating period along the $x$-axis is $\Lambda = 1.0$ µm. The incident plane-wave has wavelength $\lambda_0 = 0.499$ µm, with the polar and azimuthal incidence angles being $(\theta, \varphi)_{\text{inc}} = (150°, 0°)$. The overall diffraction efficiency is 87%, dominated by the $-2^{\text{nd}}$-order transmitted beam.

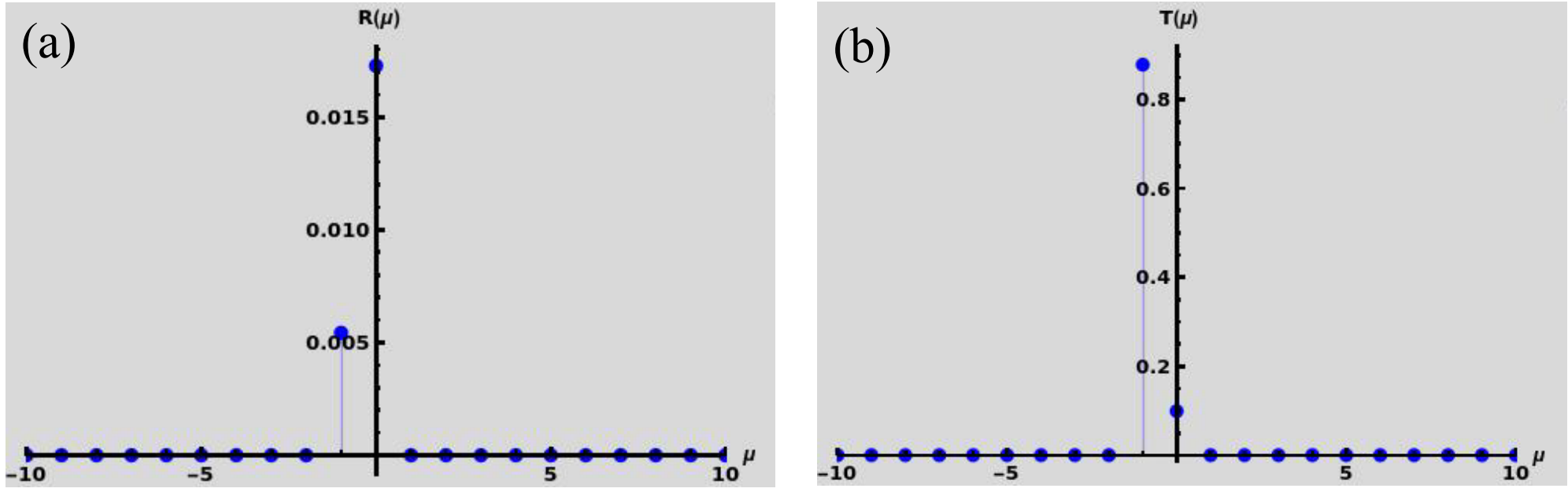

**Fig. 11**. Computed reflected and transmitted diffraction efficiencies for a 100 µm-thick grating with a period of $\Lambda = 0.50$ µm along the $x$-axis. The incident wavelength is $\lambda_0 = 0.497$ µm, and the overall diffraction efficiency is 88%.

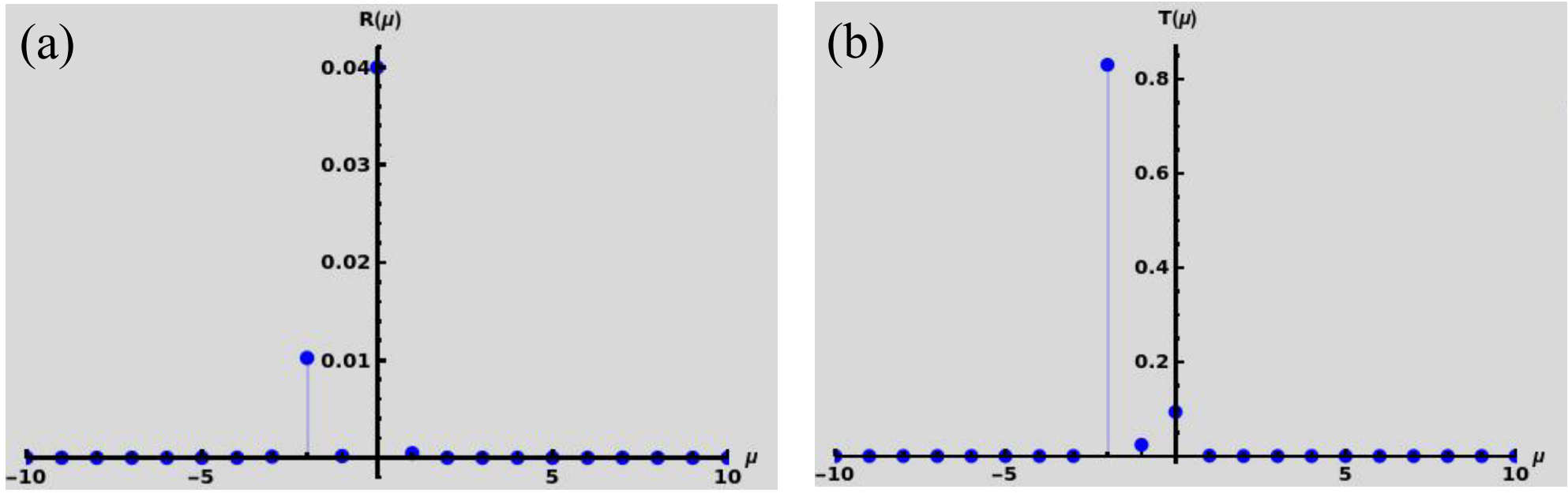

**Fig. 12**. Computed reflected and transmitted diffraction efficiencies for a 50 µm-thick grating with a period of $\Lambda = 1.0$ µm along the $x$-axis. The incident wavelength is $\lambda_0 = 0.499$ µm, and the overall diffraction efficiency is 87%, with the remaining 13% going into the $0^{\text{th}}$-order reflected and transmitted beams.

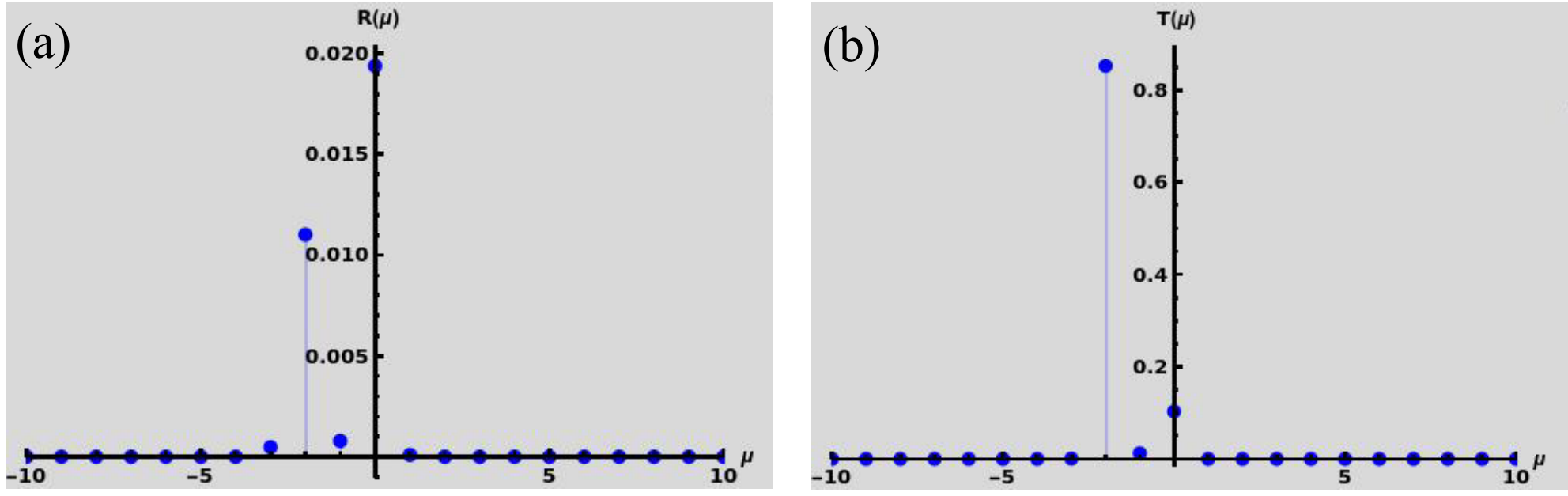

**Fig. 13**. Computed reflected and transmitted diffraction efficiencies for a 100 µm-thick grating with a period of $\Lambda = 1.5$ µm along the $x$-axis. The incident wavelength is $\lambda_0 = 0.749$ µm, and the overall diffraction efficiency is 88%, with the remaining 12% going into the $0^{\text{th}}$-order reflected and transmitted beams.

Finally, Fig.13 shows the computed reflected and transmitted diffraction efficiencies for a slab whose diagonal dielectric tensor $(\varepsilon_{h1}, \varepsilon_{h2}, \varepsilon_{h3}) = (0.07, 0.091, 0.063)$ is rotated through the Euler angles $(\theta, \varphi, \chi)_h = (45°, 30°, 36°)$. The slab's thickness is $d = 100$ µm, and its trapezoidal index-modulated profile has period $\Lambda = 1.5$ µm along the $x$-axis. The incident plane-wave has wavelength $\lambda_0 = 0.749$ µm, and the polar and azimuthal incidence angles are $(\theta, \varphi)_{\text{inc}} = (150°, 0°)$. The overall diffraction efficiency is 88%, dominated by the $-2^{\text{nd}}$-order transmitted beam.

The above results are only a small sampling from hundreds of such simulations that we carried out to confirm the conservation of energy, the stability of the numerical routines, and also the convergence of the final results when a sufficient number of coupled plane-waves were included.

**4. Rigorous coupled-wave theory of a two-dimensional birefringent holographic grating**. Figure 14 shows the basic geometry of the problem under consideration. Within the homogeneous and isotropic incidence medium having dielectric function $\varepsilon_1(\omega) = n_1^2(\omega)$, an incident plane-wave arrives at the upper surface of a holographic grating. The incident beam has $k$-vector $\boldsymbol{k}^{(\text{inc})} = (2\pi n_1/\lambda_0)(\sin\theta_0 \cos\varphi_0\, \hat{\boldsymbol{x}} + \sin\theta_0 \sin\varphi_0\, \hat{\boldsymbol{y}} + \cos\theta_0\, \hat{\boldsymbol{z}})$ and $E$-field amplitude $\boldsymbol{E}^{(\text{inc})} = E_{x0}\hat{\boldsymbol{x}} + E_{y0}\hat{\boldsymbol{y}} + E_{z0}\hat{\boldsymbol{z}}$. The beam is monochromatic, having frequency $\omega$ and vacuum wavelength $\lambda_0 = 2\pi c/\omega$. Maxwell's 1st equation, $\boldsymbol{\nabla} \cdot \boldsymbol{D}(\boldsymbol{r}, t) = 0$, imposes the constraint $\boldsymbol{k}^{(\text{inc})} \cdot \boldsymbol{E}^{(\text{inc})} = 0$ on the incident plane-wave. Also, from Maxwell's 3rd equation, $\boldsymbol{\nabla} \times \boldsymbol{E}(\boldsymbol{r}, t) = -\partial \boldsymbol{B}(\boldsymbol{r}, t)/\partial t$, we find the incident beam's $H$-field as $\boldsymbol{H}^{(\text{inc})} = (\mu_0\omega)^{-1}\boldsymbol{k}^{(\text{inc})} \times \boldsymbol{E}^{(\text{inc})}$.

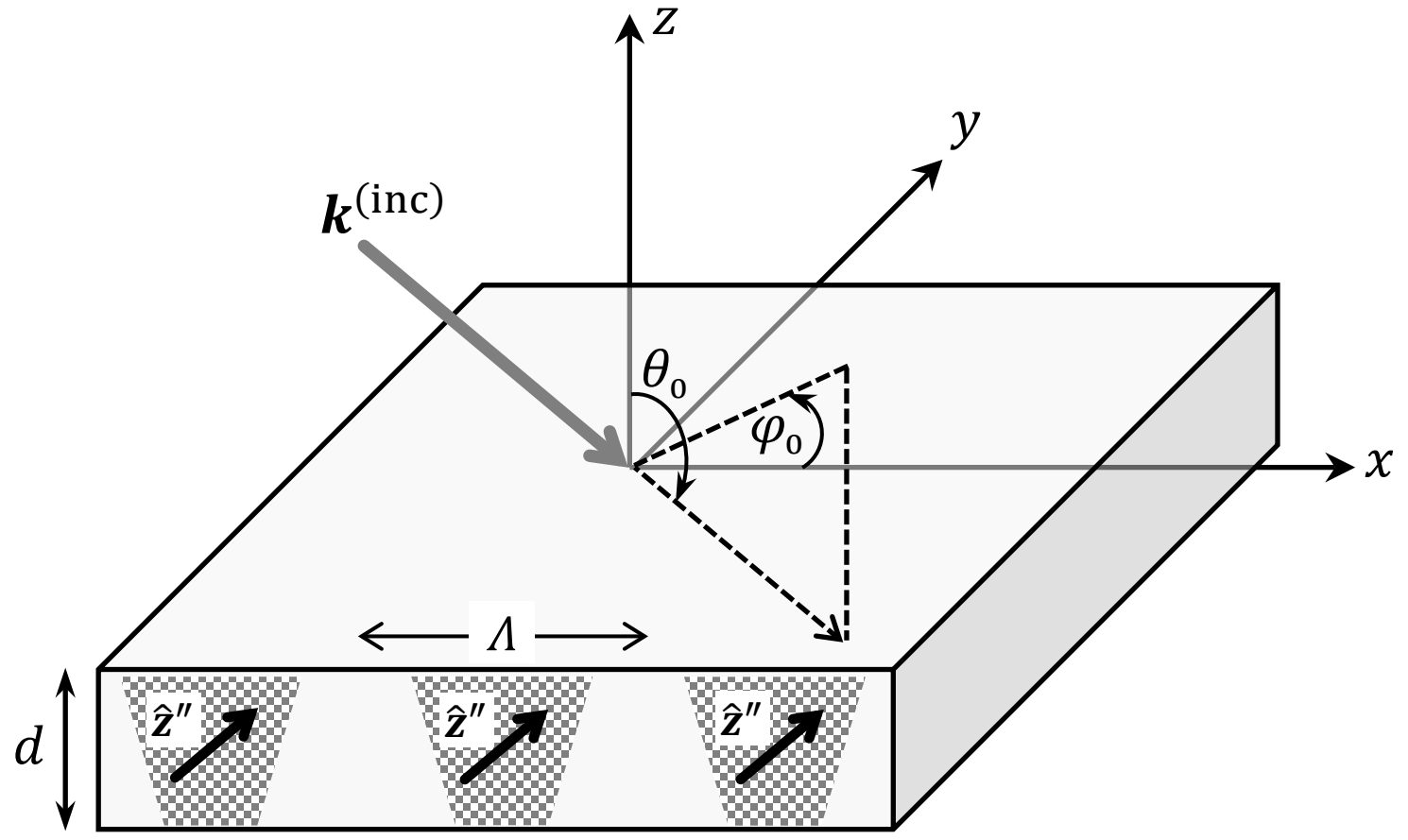

**Fig. 14**. A holographic grating of thickness $d$ and dielectric tensor $\tilde{\varepsilon}(\boldsymbol{r}, \omega) = \varepsilon_2 \tilde{I} + g(x, z)\tilde{\varepsilon}_h$ is illuminated from above by a plane-wave of frequency $\omega$, $k$-vector $\boldsymbol{k}^{(\text{inc})}$, $E$-field amplitude $\boldsymbol{E}^{(\text{inc})}$, and $H$-field amplitude $\boldsymbol{H}^{(\text{inc})}$. The index-modulation function $g(x, z)$ equals 1.0 in the shaded areas, equals 0.0 outside the shaded region, has period $\Lambda$ along the $x$-axis, and is uniformly present along the $y$-axis. The anisotropic increment $\tilde{\varepsilon}_h$ of the dielectric function within the shaded region has principal values $\varepsilon_{h1}, \varepsilon_{h2}, \varepsilon_{h3}$ along the rotated coordinate axes $x'', y'', z''$, where the Euler angles of the rotated coordinate system relative to the $xyz$ system are $\theta_h$, $\varphi_h$, and $\varphi_h - \psi_h$; see Appendix A. The dielectric functions of the incidence and transmittance media, both of which are linear, homogeneous, isotropic, transparent, and non-magnetic, are $\varepsilon_1(\omega)$ and $\varepsilon_3(\omega)$, respectively.

The holographic grating has thickness $d$ and dielectric tensor $\tilde{\varepsilon}(\boldsymbol{r}, \omega) = \varepsilon_2 \tilde{I} + g(x, z)\tilde{\varepsilon}_h$, where the periodically-repeated function $g(x, z)$ equals 1.0 in the shaded areas depicted in Fig.14, equals 0.0 outside the shaded region, has period $\Lambda$ along the $x$-axis, and is uniformly present along the $y$-axis. Within the shaded region, the Euler angles[20] of the principal axes $x'', y'', z''$ of the hologram are denoted by $\theta_h$, $\varphi_h$, and $\varphi_h - \psi_h$ (see Appendix A for a description of the Euler angles).

The base dielectric function $\varepsilon_2(\omega)$ of the hologram is augmented by a periodically-repeated modulation that is given by the tensor $g(x, z)\tilde{\varepsilon}_h = \tilde{\varepsilon}_h \sum_{m=-M}^{M} \sum_{n=-N}^{N} g_{mn} \exp[\mathrm{i}(mK_x x + nK_z z)]$. Here $K_x = 2\pi/\Lambda$ is the grating vector along the $x$-axis, while $K_z = 2\pi/d$ (with $d$ being the thickness of the holographic slab), is the "effective" grating vector along the $z$-axis. The coefficients $g_{mn}$ are the two-dimensional Fourier series coefficients of the grating function $g(x, z)$; see Eqs.(1a) and (1b). For an arbitrary $g(x,z)$, the coefficients $g_{mn}$ must be computed numerically in accordance with Eq.(1b)—with the integration interval along $z$ becoming $(-d_1, d_2)$ if one chooses to smooth out the grating profile $g(x,z)$ using the method described earlier in conjunction with Fig.6. Occasionally, however, the Fourier integral in Eq.(1b) may yield to analytic evaluation, as was the case in the Examples of Sec.2. Be it as it may, the EM fields inside the hologram can be written as follows:

$$\boldsymbol{E}(\boldsymbol{r},t) = \sum_{\mu=\mu_a}^{\mu_b} \sum_{\nu=\nu_a}^{\nu_b} \boldsymbol{E}_{\mu\nu} \exp\{\mathrm{i}[(k_{x0} + \mu K_x)x + k_{y0}y + (k_z + \nu K_z)z - \omega t]\}. \tag{8a}$$

to be determined

$$\boldsymbol{H}(\boldsymbol{r},t) = \sum_{\mu=\mu_a}^{\mu_b} \sum_{\nu=\nu_a}^{\nu_b} \boldsymbol{H}_{\mu\nu} \exp\{\mathrm{i}[(k_{x0} + \mu K_x)x + k_{y0}y + (k_z + \nu K_z)z - \omega t]\}. \tag{8b}$$

In the above equations, $k_{x0} = k_x^{(\text{inc})}$, $k_{y0} = k_y^{(\text{inc})}$, the integer index $\mu$ ranges from $\mu_a$ to $\mu_b$, and the integer index $\nu$ ranges from $\nu_a$ to $\nu_b$. The total number of plane-waves thus allowed inside the hologram is $\Gamma = (\mu_b - \mu_a + 1) \times (\nu_b - \nu_a + 1)$. The index $\kappa$ uniquely identifies each plane-wave by assigning a unique integer between 1 and $\Gamma$ to the index-pair $(\mu, \nu)$, as follows:

$$\kappa = (\mu - \mu_a + 1) + (\nu - \nu_a) \times (\mu_b - \mu_a + 1). \tag{9}$$

Table 1 lists the values of $\kappa$ assigned to each index-pair $(\mu, \nu)$ in accordance with Eq.(9). If the number of allowed values of $\mu$ are denoted by $\mathscr{k}_1 = \mu_b - \mu_a + 1$, and those of $\nu$ by $\mathscr{k}_2 = \nu_b - \nu_a + 1$, the total number of admitted plane-waves inside the hologram will be $\Gamma = \mathscr{k}_1\mathscr{k}_2$.

| $\nu \downarrow \diagdown \mu \rightarrow$ | $\mu_a$ | $\mu_a + 1$ | $\cdots$ | $\mu_b - 1$ | $\mu_b$ |
|---|---|---|---|---|---|
| $\nu_a$ | $\kappa = 1$ | $2$ | $\cdots$ | $\mathscr{k}_1 - 1$ | $\kappa = \mathscr{k}_1$ |
| $\nu_a + 1$ | $\mathscr{k}_1 + 1$ | $\mathscr{k}_1 + 2$ | $\cdots$ | $2\mathscr{k}_1 - 1$ | $2\mathscr{k}_1$ |
| $\vdots$ | $\vdots$ | $\vdots$ | $\ddots$ | $\vdots$ | $\vdots$ |
| $\nu_b - 1$ | $(\mathscr{k}_2 - 2)\mathscr{k}_1 + 1$ | $(\mathscr{k}_2 - 2)\mathscr{k}_1 + 2$ | $\cdots$ | $(\mathscr{k}_2 - 1)\mathscr{k}_1 - 1$ | $(\mathscr{k}_2 - 1)\mathscr{k}_1$ |
| $\nu_b$ | $(\mathscr{k}_2 - 1)\mathscr{k}_1 + 1$ | $(\mathscr{k}_2 - 1)\mathscr{k}_1 + 2$ | $\cdots$ | $\mathscr{k}_2\mathscr{k}_1 - 1$ | $\kappa = \mathscr{k}_2\mathscr{k}_1$ |

**Table 1**. The $\mu$ index assumes integer values between $\mu_a$ and $\mu_b$, while the $\nu$ index ranges from $\nu_a$ to $\nu_b$. Inside the hologram, admitted plane-waves are uniquely identified by their assigned integer-pair $(\mu, \nu)$, or, equivalently, by the integer index $\kappa$ defined in Eq.(9). Denoting the number of allowed values of $\mu$ by $\mathscr{k}_1 = \mu_b - \mu_a + 1$, and those of $\nu$ by $\mathscr{k}_2 = \nu_b - \nu_a + 1$, the total number of plane-waves will be $\Gamma = \mathscr{k}_1\mathscr{k}_2$.

Application of Maxwell's 3rd equation, $\boldsymbol{\nabla} \times \boldsymbol{E} = -\partial \boldsymbol{B}/\partial t$, to Eqs.(8a) and (8b) now yields

$$[(k_{x0} + \mu K_x)\hat{\boldsymbol{x}} + k_{y0}\hat{\boldsymbol{y}} + (k_z + \nu K_z)\hat{\boldsymbol{z}}] \times (E_{x,\mu\nu}\hat{\boldsymbol{x}} + E_{y,\mu\nu}\hat{\boldsymbol{y}} + E_{z,\mu\nu}\hat{\boldsymbol{z}})$$
$$= \omega\mu_0(H_{x,\mu\nu}\hat{\boldsymbol{x}} + H_{y,\mu\nu}\hat{\boldsymbol{y}} + H_{z,\mu\nu}\hat{\boldsymbol{z}}). \tag{10}$$

The individual $x, y, z$ components of the above equation are written as follows:

$$\mu_0\omega H_{x,\mu\nu} = k_{y0}E_{z,\mu\nu} - (k_z + \nu K_z)E_{y,\mu\nu}, \tag{10a}$$

$$\mu_0\omega H_{y,\mu\nu} = (k_z + \nu K_z)E_{x,\mu\nu} - (k_{x0} + \mu K_x)E_{z,\mu\nu}, \tag{10b}$$

$$\mu_0\omega H_{z,\mu\nu} = (k_{x0} + \mu K_x)E_{y,\mu\nu} - k_{y0}E_{x,\mu\nu}. \tag{10c}$$

Equations (10a) and (10b) are not useful yet, as they contain the $z$-component $E_{z,\mu\nu}$ of the $E$-field; this field component will be determined shortly. As for Eq.(10c), it will be incorporated into the remaining Maxwell equation in order to eliminate the $z$-component $H_{z,\mu\nu}$ of the $H$-field. Needless to say, Eq.(10) guarantees the satisfaction of Maxwell's 4$^{\text{th}}$ equation, $\boldsymbol{\nabla}\cdot\boldsymbol{B}(\boldsymbol{r},t)=0$, since $\boldsymbol{\nabla}\cdot(\boldsymbol{\nabla}\times\boldsymbol{E})$ is automatically equal to zero, while $\boldsymbol{\nabla}\cdot(\partial\boldsymbol{B}/\partial t)=-\mathrm{i}\omega\boldsymbol{\nabla}\cdot\boldsymbol{B}$. We now turn to Maxwell's 2$^{\text{nd}}$ equation, $\boldsymbol{\nabla}\times\boldsymbol{H}=\partial\boldsymbol{D}/\partial t$, and use Eqs.(8a) and (8b) to arrive at

$$\sum_{\mu_a}^{\mu_b}\sum_{\nu_a}^{\nu_b}[(k_{x0}+\mu K_x)\hat{\boldsymbol{x}}+k_{y0}\hat{\boldsymbol{y}}+(k_z+\nu K_z)\hat{\boldsymbol{z}}]\times(H_{x,\mu\nu}\hat{\boldsymbol{x}}+H_{y,\mu\nu}\hat{\boldsymbol{y}}+H_{z,\mu\nu}\hat{\boldsymbol{z}})$$
$$\times\exp\{\mathrm{i}[(k_{x0}+\mu K_x)x+k_{y0}y+(k_z+\nu K_z)z-\omega t]\}$$
$$=-\omega\varepsilon_0\{\varepsilon_2\tilde{I}+\tilde{\varepsilon}_h\textstyle\sum_{m=-M}^{M}\sum_{n=-N}^{N}g_{mn}\exp[\mathrm{i}(mK_xx+nK_zz)]\}$$
$$\times\sum_{\mu_a}^{\mu_b}\sum_{\nu_a}^{\nu_b}\boldsymbol{E}_{\mu\nu}\exp\{\mathrm{i}[(k_{x0}+\mu K_x)x+k_{y0}y+(k_z+\nu K_z)z-\omega t]\}. \quad (11)$$

For each integer-pair $(\mu,\nu)$, or, equivalently, for plane-waves with the corresponding $\kappa$ indices, the $x,y,z$ components of the above equation now yield the following coupled-wave equations:

$$k_{y0}H_{z,\mu\nu}-(k_z+\nu K_z)H_{y,\mu\nu}=-\varepsilon_0\varepsilon_2\omega E_{x,\mu\nu}$$
$$-\varepsilon_0\omega\textstyle\sum_m\sum_n g_{mn}(\varepsilon_{xx}E_{x,\mu-m,\nu-n}+\varepsilon_{xy}E_{y,\mu-m,\nu-n}+\varepsilon_{xz}E_{z,\mu-m,\nu-n}), \quad (11a)$$

$$(k_z+\nu K_z)H_{x,\mu\nu}-(k_{x0}+\mu K_x)H_{z,\mu\nu}=-\varepsilon_0\varepsilon_2\omega E_{y,\mu\nu}$$
$$-\varepsilon_0\omega\textstyle\sum_m\sum_n g_{mn}(\varepsilon_{yx}E_{x,\mu-m,\nu-n}+\varepsilon_{yy}E_{y,\mu-m,\nu-n}+\varepsilon_{yz}E_{z,\mu-m,\nu-n}), \quad (11b)$$

$$(k_{x0}+\mu K_x)H_{y,\mu\nu}-k_{y0}H_{x,\mu\nu}=-\varepsilon_0\varepsilon_2\omega E_{z,\mu\nu}$$
$$-\varepsilon_0\omega\textstyle\sum_m\sum_n g_{mn}(\varepsilon_{zx}E_{x,\mu-m,\nu-n}+\varepsilon_{zy}E_{y,\mu-m,\nu-n}+\varepsilon_{zz}E_{z,\mu-m,\nu-n}). \quad (11c)$$

We mention in passing that Eq.(11) guarantees that Maxwell's 1$^{\text{st}}$ equation, $\boldsymbol{\nabla}\cdot\boldsymbol{D}(\boldsymbol{r},t)=0$, is automatically satisfied. This is because $\boldsymbol{\nabla}\cdot(\boldsymbol{\nabla}\times\boldsymbol{H})$ vanishes, while $\boldsymbol{\nabla}\cdot(\partial\boldsymbol{D}/\partial t)=-\mathrm{i}\omega\boldsymbol{\nabla}\cdot\boldsymbol{D}$.

Whereas Eqs.(11a) and (11b) contain the as-yet-unknown parameter $k_z$, the last of the above equations, Eq.(11c), is free from this unknown parameter. One may thus relate, via Eq.(11c), the $z$-components $E_{z,\mu\nu}$ of the various plane-waves to the remaining components $(E_x,E_y,H_x,H_y)_{\mu\nu}$ of *all* admitted plane-waves—namely, those having $\mu_a\le\mu\le\mu_b$ and $\nu_a\le\nu\le\nu_b$. Each admitted plane-wave is, of course, identified with a unique integer index $\kappa$ between 1 and $\Gamma$, in accordance with Eq.(9). To ensure dimensional compatibility among the various parameters and coefficients, we shall henceforth multiply the $H$-field components with the impedance of free space, $Z_0=(\mu_0/\varepsilon_0)^{½}\cong 376.730313\ \Omega$. In matrix form, Eq.(11c) is now written as follows:

$$\underbrace{\begin{pmatrix}P_{1,1}&\cdots&P_{1,\kappa}&\cdots&P_{1,\Gamma}\\ \vdots&\ddots&\vdots&\ddots&\vdots\\ P_{\kappa,1}&\cdots&P_{\kappa,\kappa}&\cdots&P_{\kappa,\Gamma}\\ \vdots&\ddots&\vdots&\ddots&\vdots\\ P_{\Gamma,1}&\cdots&P_{\Gamma,\kappa}&\cdots&P_{\Gamma,\Gamma}\end{pmatrix}}_{\Gamma\times\Gamma}\begin{pmatrix}E_{z,1}\\ \vdots\\ E_{z,\kappa}\\ \vdots\\ E_{z,\Gamma}\end{pmatrix}=\left(\underbrace{\begin{matrix}Q_{1,1}&\cdots&Q_{1,2\Gamma}\\ Q_{2,1}&\cdots&Q_{2,2\Gamma}\\ \vdots&\ddots&\vdots\\ Q_{\kappa,1}&\cdots&Q_{\kappa,2\Gamma}\\ \vdots&\ddots&\vdots\\ Q_{\Gamma,1}&\cdots&Q_{\Gamma,2\Gamma}\end{matrix}}_{\Gamma\times2\Gamma}\ \middle|\ \underbrace{\begin{matrix}R_{1,1}&R_{1,2}&0&0&\cdots&0\\ 0&0&R_{2,3}&R_{2,4}&0&0\\ 0&0&0&0&\ddots&0\\ 0&0&0&\ddots&\ddots&0\\ \vdots&\vdots&\vdots&\ddots&\ddots&\vdots\\ 0&0&0&0&R_{\Gamma,2\Gamma-1}&R_{\Gamma,2\Gamma}\end{matrix}}_{\Gamma\times2\Gamma}\right)\begin{pmatrix}\vdots\\ E_{x,\kappa}\\ E_{y,\kappa}\\ \vdots\\ Z_0H_{x,\kappa}\\ Z_0H_{y,\kappa}\\ \vdots\end{pmatrix}. \quad (12)$$

To form the square $(\Gamma \times \Gamma)$ matrix $P$, set its diagonal elements to $\varepsilon_2 + g_{00}\varepsilon_{zz}$. Then, for each remaining element $P_{\kappa_1,\kappa_2}$, identify the pairs of integers $(\mu_1, \nu_1)$ and $(\mu_2, \nu_2)$, which are associated, respectively, with $\kappa_1$ and $\kappa_2$, as shown in Table 1, and set $P_{\kappa_1,\kappa_2} = g_{(\mu_1-\mu_2),(\nu_1-\nu_2)}\varepsilon_{zz}$.

To compute the elements $Q_{\kappa_1,\kappa_2}$ of the $\Gamma \times 2\Gamma$ matrix $Q$, use Table 1 to identify the integer pair $(\mu_1, \nu_1)$ associated with $\kappa_1$. As for $\kappa_2$, if it is even, set $\kappa_2' = \kappa_2/2$, but if it is odd, set $\kappa_2' = (\kappa_2 + 1)/2$. Use Table 1 to identify the integer pair $(\mu_2, \nu_2)$ associated with $\kappa_2'$. When $\kappa_2$ is odd, set $Q_{\kappa_1,\kappa_2} = -g_{(\mu_1-\mu_2),(\nu_1-\nu_2)}\varepsilon_{zx}$, and when $\kappa_2$ is even, set $Q_{\kappa_1,\kappa_2} = -g_{(\mu_1-\mu_2),(\nu_1-\nu_2)}\varepsilon_{zy}$.

The $\Gamma \times 2\Gamma$ matrix $R$ has only two non-zero elements in each of its rows. These non-zero elements are $R_{\kappa,2\kappa-1} = ck_{y0}/\omega$ and $R_{\kappa,2\kappa} = -c(k_{x0} + \mu K_x)/\omega$. Here the constant $c$ is the speed of light in vacuum, namely, $c = (\mu_0\varepsilon_0)^{-½} = 299{,}792{,}458$ m/sec. Needless to say, the row number $\kappa$ determines the integer pair $(\mu, \nu)$ via Table 1. Whereas the index $\mu$ appears in the preceding expression for $R_{\kappa,2\kappa}$, the corresponding index $\nu$ does *not* contribute to the $R$ matrix.

The matrix $P$ must now be inverted and multiplied into the $\Gamma \times 4\Gamma$ matrix $[Q|R]$ to yield a new $\Gamma \times 4\Gamma$ matrix $S = P^{-1}[Q|R]$, which relates the column vector $[E_{z,1}, \cdots, E_{z,\kappa}, \cdots, E_{z,\Gamma}]^T$ to the column vector $[E_{x,1}, E_{y,1}, \cdots, E_{x,\Gamma}, E_{y,\Gamma}, Z_0H_{x,1}, Z_0H_{y,1}, \cdots, Z_0H_{x,\Gamma}, Z_0H_{y,\Gamma}]^T$. At this point we have arrived at the matrix $S$, which relates the $E_z$ values of each and every plane-wave inside the hologram to the $E_x, E_y, H_x, H_y$ values associated with *all* the plane-waves inside the hologram.

Next, we rearrange Eqs.(10a) and (10b), and also Eqs.(11a) and (11b), in such a way as to enable a determination of acceptable values of $k_z$ as the eigenvalues of a large $(4\Gamma \times 4\Gamma)$ matrix. Recalling that $H_{z,\mu\nu}$ is given by Eq.(10c), and that $E_{z,\mu\nu}$ is the solution of Eq.(12), we will have

$$(k_{x0} + \mu K_x)E_{z,\mu\nu} - \nu K_z E_{x,\mu\nu} + (\omega/c)Z_0 H_{y,\mu\nu} = k_z E_{x,\mu\nu}. \quad (13a)$$

$$k_{y0}E_{z,\mu\nu} - \nu K_z E_{y,\mu\nu} - (\omega/c)Z_0 H_{x,\mu\nu} = k_z E_{y,\mu\nu}. \quad (13b)$$

$$-(\omega/c)\textstyle\sum_m\sum_n g_{mn}(\varepsilon_{yx}E_{x,\mu-m,\nu-n} + \varepsilon_{yy}E_{y,\mu-m,\nu-n} + \varepsilon_{yz}E_{z,\mu-m,\nu-n}) - (k_{x0} + \mu K_x)(ck_{y0}/\omega)E_{x,\mu\nu}$$
$$+[(c/\omega)(k_{x0} + \mu K_x)^2 - (\omega/c)\varepsilon_2]E_{y,\mu\nu} - \nu K_z Z_0 H_{x,\mu\nu} = k_z Z_0 H_{x,\mu\nu}. \quad (13c)$$

$$(\omega/c)\textstyle\sum_m\sum_n g_{mn}(\varepsilon_{xx}E_{x,\mu-m,\nu-n} + \varepsilon_{xy}E_{y,\mu-m,\nu-n} + \varepsilon_{xz}E_{z,\mu-m,\nu-n}) + [(\omega/c)\varepsilon_2 - (c/\omega)k_{y0}^2]E_{x,\mu\nu}$$
$$+(k_{x0} + \mu K_x)(ck_{y0}/\omega)E_{y,\mu\nu} - \nu K_z Z_0 H_{y,\mu\nu} = k_z Z_0 H_{y,\mu\nu}. \quad (13d)$$

Finally, we are in a position to write Eqs.(13) in the form of an eigenvalue-eigenvector equation involving a large $(4\Gamma \times 4\Gamma)$ matrix $A$, as follows:

$$\left(\begin{array}{ccc|ccc} A_{1,1} & \cdots & A_{1,2\Gamma} & A_{1,2\Gamma+1} & \cdots & A_{1,4\Gamma} \\ \vdots & \ddots & \vdots & \vdots & \ddots & \vdots \\ A_{2\Gamma,1} & \cdots & A_{2\Gamma,2\Gamma} & A_{2\Gamma,2\Gamma+1} & \cdots & A_{2\Gamma,4\Gamma} \\ \hline A_{2\Gamma+1,1} & \cdots & A_{2\Gamma+1,2\Gamma} & A_{2\Gamma+1,2\Gamma+1} & \cdots & A_{2\Gamma+1,4\Gamma} \\ \vdots & \ddots & \vdots & \vdots & \ddots & \vdots \\ A_{4\Gamma,1} & \cdots & A_{4\Gamma,2\Gamma} & A_{4\Gamma,2\Gamma+1} & \cdots & A_{4\Gamma,4\Gamma} \end{array}\right) \begin{pmatrix} \vdots \\ E_{x,\kappa} \\ E_{y,\kappa} \\ \vdots \\ Z_0H_{x,\kappa} \\ Z_0H_{y,\kappa} \\ \vdots \end{pmatrix} = k_z \begin{pmatrix} \vdots \\ E_{x,\kappa} \\ E_{y,\kappa} \\ \vdots \\ Z_0H_{x,\kappa} \\ Z_0H_{y,\kappa} \\ \vdots \end{pmatrix}. \quad (14)$$

To populate the matrix $A$, first set all its elements to zero, then build up the matrix using the following algorithm:

**1)** Use Eq.(13a) to populate rows $1, 3, 5, 7, \cdots, (2\Gamma - 1)$ of matrix $A$. For each integer pair $(\mu, \nu)$, find the index $\kappa$ of the plane-wave from Eq.(9), then designate $2\kappa - 1$ as the corresponding row-number for matrix $A$. For *all* values of the index $\ell$ (i.e., $\ell$ ranging from 1 to 4Γ), set the initial values of $A_{(2\kappa-1),\ell}$ equal to $(k_{x0} + \mu K_x) S_{\kappa,\ell}$. Increment the element $A_{(2\kappa-1),(2\kappa-1)}$ by $-\nu K_z$. Finally, increment the element $A_{(2\kappa-1),(2\Gamma+2\kappa)}$ by $\omega/c$.

**2)** Use Eq.(13b) to populate rows $2, 4, 6, 8, \cdots, 2\Gamma$ of matrix $A$. For each integer pair $(\mu, \nu)$, find the index $\kappa$ of the plane-wave from Eq.(9), then designate $2\kappa$ as the corresponding row-number for matrix $A$. For *all* values of the index $\ell$ (ranging from 1 to 4Γ), set the initial values of $A_{2\kappa,\ell}$ equal to $k_{y0} S_{\kappa,\ell}$. Increment the element $A_{2\kappa,2\kappa}$ by $-\nu K_z$ and the element $A_{2\kappa,(2\Gamma+2\kappa-1)}$ by $-\omega/c$.

**3)** Use Eq.(13c) to populate rows $(2\Gamma + 1), (2\Gamma + 3), \cdots, (4\Gamma - 1)$ of matrix $A$. For each integer pair $(\mu, \nu)$, use Eq.(9) to find the index $\kappa$ of the plane-wave, then designate $2\Gamma + 2\kappa - 1$ as the corresponding row-number for matrix $A$.

**Step 1**: Set $A_{(2\Gamma+2\kappa-1),(2\kappa-1)}$ to $-(k_{x0} + \mu K_x)(c k_{y0}/\omega)$.

**Step 2**: Set $A_{(2\Gamma+2\kappa-1),2\kappa}$ to $(c/\omega)(k_{x0} + \mu K_x)^2 - (\omega/c)\varepsilon_2$.

**Step 3**: Set $A_{(2\Gamma+2\kappa-1),(2\Gamma+2\kappa-1)}$ to $-\nu K_z$.

**Step 4**: Choose an integer pair $(m, n)$ and find the index $\kappa'$ of the plane-wave $(\mu - m, \nu - n)$.

**Step 5**: Increment $A_{(2\Gamma+2\kappa-1),(2\kappa'-1)}$ by $-(\omega/c) g_{mn} \varepsilon_{yx}$.

**Step 6**: Increment $A_{(2\Gamma+2\kappa-1),2\kappa'}$ by $-(\omega/c) g_{mn} \varepsilon_{yy}$.

**Step 7**: For *all* indices $\ell$ (ranging from 1 to 4Γ), increment $A_{(2\Gamma+2\kappa-1),\ell}$ by $-(\omega/c) g_{mn} \varepsilon_{yz} S_{\kappa',\ell}$.

**Step 8**: Repeat Steps 4-7 until all allowed values of $(m, n)$ are exhausted.

**4)** Use Eq.(13d) to populate rows $(2\Gamma + 2), (2\Gamma + 4), \cdots, 4\Gamma$ of matrix $A$. For each integer pair $(\mu, \nu)$, use Eq.(9) to find the index $\kappa$ of the plane-wave, then designate $2\Gamma + 2\kappa$ as the corresponding row-number for matrix $A$.

**Step 1**: Set $A_{(2\Gamma+2\kappa),(2\kappa-1)}$ to $(\omega/c)\varepsilon_2 - (c/\omega) k_{y0}^2$.

**Step 2**: Set $A_{(2\Gamma+2\kappa),2\kappa}$ to $(k_{x0} + \mu K_x)(c k_{y0}/\omega)$.

**Step 3**: Set $A_{(2\Gamma+2\kappa),(2\Gamma+2\kappa)}$ to $-\nu K_z$.

**Step 4**: Choose an integer pair $(m, n)$ and find the index $\kappa'$ of the plane-wave $(\mu - m, \nu - n)$.

**Step 5**: Increment $A_{(2\Gamma+2\kappa),(2\kappa'-1)}$ by $(\omega/c) g_{mn} \varepsilon_{xx}$.

**Step 6**: Increment $A_{(2\Gamma+2\kappa),2\kappa'}$ by $(\omega/c) g_{mn} \varepsilon_{xy}$.

**Step 7**: For *all* indices $\ell$ (ranging from 1 to 4Γ), increment $A_{(2\Gamma+2\kappa),\ell}$ by $(\omega/c) g_{mn} \varepsilon_{xz} S_{\kappa',\ell}$.

**Step 8**: Repeat Steps 4-7 until all allowed values of $(m, n)$ are exhausted.

Once the matrix $A$ is determined, one proceeds to find its eigenvalues and eigenvectors, then use these values to match the boundary conditions at the top and bottom facets of the holographic slab. A detailed description of this matching of the boundary conditions is given in Appendix C.

The reflected and transmitted plane-waves will have $(k_x, k_y) = (k_{x0} + \mu K_x, k_{y0})$, for a total of $\mathcal{k}_1 = \mu_b - \mu_a + 1$ plane-waves in each medium. Each of these plane-waves, however, must have a $\mathcal{k}_2$-fold degeneracy, where $\mathcal{k}_2 = \nu_b - \nu_a + 1$ is the number of the Fourier coefficients used in Eq.(11) to account for the $z$-dependence of the function $g(x, z)$. Accounting for these degeneracies thus leads to a total number $\Gamma = \mathcal{k}_1 \mathcal{k}_2$ of reflected plane-waves in medium 1, and also the same number of transmitted plane-waves in medium 3. The total number of unknowns (including one scale-factor for each eigenvector of the matrix $A$) is, therefore, $8\Gamma$, which dictates the inversion of an $8\Gamma \times 8\Gamma$ matrix in order to determine all the unknown coefficients. The unknowns include $2\Gamma$ values for $E_{x,\mu}^{(\text{ref})}$ and $E_{y,\mu}^{(\text{ref})}$ associated with the reflected planes-waves, another $2\Gamma$ values for the $E_x^{(\text{trans})}$ and $E_y^{(\text{trans})}$ of the transmitted planes-waves, and $4\Gamma$ values $(\zeta_1, \zeta_2, \cdots, \zeta_{4\Gamma})$ for the scale-factors associated with each eigenvector of the matrix $A$ of Eq.(14). Note that each reflected (and also each transmitted) plane-wave has a $\mathcal{k}_2$-fold degeneracy, which requires that, in the end, we add up the $\mathcal{k}_2$ field amplitudes obtained for each and every one of the diffracted orders, of which there are $\mathcal{k}_1$ in the incidence medium 1, and also $\mathcal{k}_1$ in the transmittance medium 3.

An important question arises as to how one should treat the incident beam, whose $E$-field amplitudes $E_x^{(\text{inc})}$ and $E_y^{(\text{inc})}$ contribute to the boundary conditions at the interface between the incidence medium 1 and the holographic medium 2. It turns out that a substantial degree of flexibility is provided by the $\mathcal{k}_2$-fold degeneracy of the $0^{\text{th}}$ order reflected beam, whose tangential $E$- and $H$-field components must combine with those of the incident beam in order to match the (superposed) tangential components of the corresponding $\mathcal{k}_2$ diffracted orders $(\mu = 0, \nu_a \le \nu \le \nu_b)$ inside the hologram. Thus, the incident beam may be paired with *any* one of the degenerate $0^{\text{th}}$ order reflected plane-waves—i.e., any one out of $\mathcal{k}_2$ reflected plane-waves which have $\mu = 0$, or, equivalently, $k_x^{(\text{ref})} = k_x^{(\text{inc})} = k_{x0}$ and $k_y^{(\text{ref})} = k_y^{(\text{inc})} = k_{y0}$. This implies that all the remaining $0^{\text{th}}$-order reflected beams must be paired with a non-existent incident beam, that is, one for which $E_x^{(\text{inc})} = E_y^{(\text{inc})} = 0$. Alternatively, one may choose to split the incident beam (in any arbitrary way) among two or more $0^{\text{th}}$-order reflected beams. The final results of the calculations should *not* depend on the specific way in which the incident beam has been shared among the various (degenerate) $0^{\text{th}}$-order reflected beams.

**5. Concluding remarks**. It may be argued, for the following two reasons, that the adjective "rigorous" used in conjunction with the RCWA method is misleading:

i) The Fourier series expansion of the hologram's refractive index (or dielectric tensor) may or may not have a finite number of Fourier components. In the latter case, when the spatial frequencies of $\tilde{\varepsilon}(x, z)$ extend all the way to infinity, the Fourier series must, of necessity, be truncated at a reasonably large frequency, thus leading to some level of inaccuracy in the numerical results.

ii) Even in special cases where the Fourier series expansion of $\tilde{\varepsilon}(x, z)$ happens to contain a finite number of elements, one is obligated to limit the number of coupled plane-waves within the holographic slab (and also those outside the slab) to some finite value, which is a form of approximation since, by doing so, one eliminates the contributions of high-spatial-frequency waves to the EM fields inside (as well as outside) the hologram.

These are valid concerns, of course, but, in practice, one truncates the Fourier series of $\tilde{\varepsilon}(x, z)$ at reasonably large spatial frequencies and, similarly, limits the computation to a reasonable number of coupled plane-waves. One then proceeds to check and see if inclusion of additional terms would impact the final numerical results of the computation. Once the desired accuracy in the final results has been achieved, there will be no point in including higher spatial frequencies in the calculations.

For practical applications of the proposed algorithm, the computation speed is of utmost significance, and one might inquire as to whether the proposed method is superior in this regard to the conventional method of slicing thick holograms — or, for that matter, to alternative methods of treating such diffraction gratings[21-24]. The answer would depend on several factors, e.g., the number of slices that would be necessary for an accurate representation of the $z$-dependence of the dielectric tensor. It will also depend on the parallel-processing capabilities of the computer that will be used, and the degree to which respective algorithms could benefit from parallelization of their numerics. A potential advantage of our proposed method vis-à-vis the slicing method is that, in addition to the reflected and transmitted diffraction efficiencies, our method generates — as a readily available by-product of its computations — the EM field distribution inside the hologram. In contrast, the artificial layering used in the slicing method could produce enough spurious scatterings at the boundaries between adjacent layers to render the computed EM field inside a thick hologram noisy and (potentially) unreliable.

Finally, the discussions of the present paper have been restricted to holograms whose refractive indices (or dielectric tensors) are two-dimensional functions of $x$ and $z$ — with periodic modulation along the $x$-axis. However, there are no restrictions, in principle, to these two dimensions; in other words, the proposed algorithm can be readily extended to thick holograms whose index-modulation has periodicity along both the $x$ and $y$ axes, in addition to having a nonuniform distribution in the thickness direction (i.e., along the $z$-axis). The method may also be further generalized to allow for the inclusion of optical activity (in addition to birefringence) in the hologram's dielectric tensor.

## Appendix A

## Dielectric Tensor in a Rotated Coordinate System

Our holographic grating slab has thickness $d$ and dielectric tensor $\tilde{\varepsilon}(\boldsymbol{r},\omega) = \varepsilon_2 \tilde{I} + g(x,z)\tilde{\varepsilon}_h$, where the periodically-repeated function $g(x,z)$ has period $\Lambda$ along the $x$-axis, and is uniformly present along the $y$-axis. Within the rectangular block of the unit-cell (i.e., where $0 \le x < \Lambda$ and $-d \le z \le 0$), the function $g(x,z)$ is zero where the refractive index is simply $n_2 = \sqrt{\varepsilon_2}$, and equals 1.0 where the base dielectric tensor $\varepsilon_2 \tilde{I}$ is augmented by $\tilde{\varepsilon}_h$. In the region of the unit-cell where $g(x,z) = 1$, the Euler angles[20] of the principal axes $x'', y'', z''$ of the hologram are denoted by $\theta_h$, $\varphi_h$, and $\varphi_h - \psi_h$.

To understand the Euler angles, take the standard Cartesian $xyz$ coordinate system and rotate it around the $z$-axis through the angle $\psi_h$ (measured from $x$ toward $y$) to arrive at the $x'y'z'$ system as depicted in Fig. A1(a). Imagine the $x'y'z'$ axes as forming a solid tripod, then hold onto the $z'$-axis (which is the same as $z$ at this point) and rotate the entire $x'y'z'$ tripod through the angle $\theta_h$ while keeping the azimuthal angle $\varphi$ (as seen in the $xyz$ system) constant at $\varphi_h$; see Fig. A1(b). The new coordinate system is the $x''y''z''$ system of the principal refractive indices. In the $x''y''z''$ system the modulated dielectric constants of the hologram are $\varepsilon_{h1}$, $\varepsilon_{h2}$, $\varepsilon_{h3}$, where the subscripts $1, 2, 3$ refer to the directions along $x'', y'', z''$, respectively.

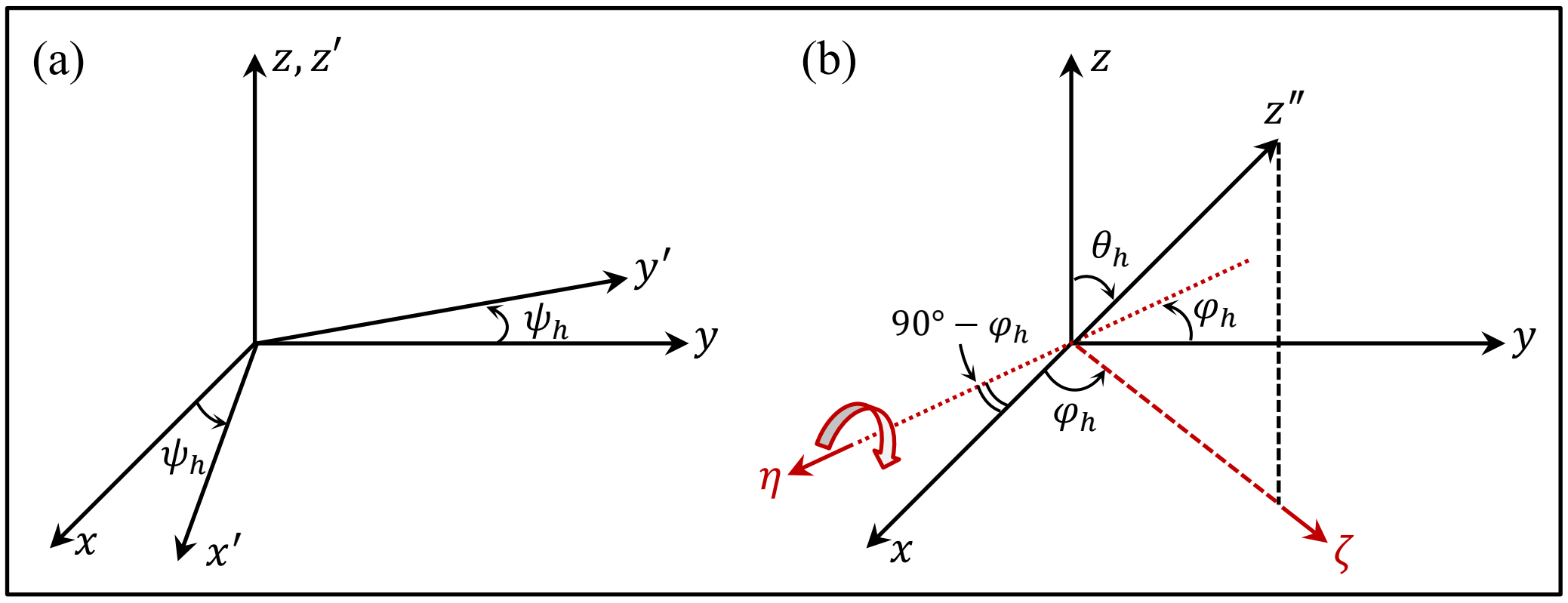

**Fig. A1**. Description of the Euler angles $\theta_h$, $\varphi_h$, and $\varphi_h - \psi_h$. (a) The $xyz$ coordinate system is first rotated around the $z$-axis through the angle $\psi_h$, producing the rotated coordinate system $x'y'z'$. (b) The new system is tilted (as if it were a rigid tripod) in such a way that its $z'$-axis turns away from the $z$-axis through the angle $\theta_h$, while maintaining a constant azimuthal angle $\varphi_h$. This brings the $x', y', z'$ axes to their final positions, namely, $x'', y'', z''$, respectively. (To reduce clutter, the $x''$ and $y''$ axes are not shown.) The dotted line, which is the original $y$-axis rotated around $z$ through the angle $\varphi_h$, is the rotation axis (within the $xy$-plane) around which the entire $x'y'z'$ system rotates to become (or to coincide with) the $x''y''z''$ system. This rotation axis is identified as the $\eta$-axis, which, together with an orthogonal axis $\zeta$ (also located in the $xy$-plane), forms an intermediate $\eta\zeta$ pair of axes within the $xy$-plane. (Note that any arbitrarily-oriented $x''y''z''$ system can be brought into alignment with $xyz$ by performing these two rotations in reverse order.)

Let us now consider an electric field $\boldsymbol{E} = E_x\hat{\boldsymbol{x}} + E_y\hat{\boldsymbol{y}} + E_z\hat{\boldsymbol{z}}$ specified within the $xyz$ coordinate system. The projections of this field onto the $\eta$ and $\zeta$ axes, which are the intermediate axes depicted in Fig. A1(b), are readily seen to be

$$E_\eta = E_x \sin\varphi_h - E_y \cos\varphi_h, \tag{A1a}$$

$$E_\zeta = E_x \cos\varphi_h + E_y \sin\varphi_h. \tag{A1b}$$

The angle between the $x'$-axis and the rotation axis $\eta$ is $½\pi - \varphi_h + \psi_h$, while that between $y'$ and $\eta$ is $\varphi_h - \psi_h$. These angles do not change when the $x'y'z'$ system rotates around $\eta$ through the angle $\theta_h$. We thus have

$$\begin{aligned} E_{x''} = {} & E_x \sin\varphi_h \sin(\varphi_h - \psi_h) + E_x \cos\varphi_h \cos\theta_h \cos(\varphi_h - \psi_h) \\ & -E_y \cos\varphi_h \sin(\varphi_h - \psi_h) + E_y \sin\varphi_h \cos\theta_h \cos(\varphi_h - \psi_h) \\ & -E_z \sin\theta_h \cos(\varphi_h - \psi_h), \end{aligned} \tag{A2a}$$

$$\begin{aligned} E_{y''} = {} & -E_x \sin\varphi_h \cos(\varphi_h - \psi_h) + E_x \cos\varphi_h \cos\theta_h \sin(\varphi_h - \psi_h) \\ & +E_y \cos\varphi_h \cos(\varphi_h - \psi_h) + E_y \sin\varphi_h \cos\theta_h \sin(\varphi_h - \psi_h) \\ & -E_z \sin\theta_h \sin(\varphi_h - \psi_h), \end{aligned} \tag{A2b}$$

$$E_{z''} = E_x \cos\varphi_h \sin\theta_h + E_y \sin\varphi_h \sin\theta_h + E_z \cos\theta_h. \tag{A2c}$$

In matrix form, the above equations are summarized as

$$\begin{pmatrix} E_{x''} \\ E_{y''} \\ E_{z''} \end{pmatrix} = \underbrace{\begin{pmatrix} \eta_{xx} & \eta_{xy} & \eta_{xz} \\ \eta_{yx} & \eta_{yy} & \eta_{yz} \\ \eta_{zx} & \eta_{zy} & \eta_{zz} \end{pmatrix}}_{\tilde{\eta}} \begin{pmatrix} E_x \\ E_y \\ E_z \end{pmatrix}, \tag{A3}$$

where

$$\eta_{xx} = \cos\theta_h \cos\varphi_h \cos(\varphi_h - \psi_h) + \sin\varphi_h \sin(\varphi_h - \psi_h). \tag{A3a}$$

$$\eta_{xy} = \cos\theta_h \sin\varphi_h \cos(\varphi_h - \psi_h) - \cos\varphi_h \sin(\varphi_h - \psi_h). \tag{A3b}$$

$$\eta_{xz} = -\sin\theta_h \cos(\varphi_h - \psi_h). \tag{A3c}$$

$$\eta_{yx} = \cos\theta_h \cos\varphi_h \sin(\varphi_h - \psi_h) - \sin\varphi_h \cos(\varphi_h - \psi_h). \tag{A3d}$$

$$\eta_{yy} = \cos\theta_h \sin\varphi_h \sin(\varphi_h - \psi_h) + \cos\varphi_h \cos(\varphi_h - \psi_h). \tag{A3e}$$

$$\eta_{yz} = -\sin\theta_h \sin(\varphi_h - \psi_h). \tag{A3f}$$

$$\eta_{zx} = \sin\theta_h \cos\varphi_h. \tag{A3g}$$

$$\eta_{zy} = \sin\theta_h \sin\varphi_h. \tag{A3h}$$

$$\eta_{zz} = \cos\theta_h. \tag{A3i}$$

Thus, in the $x''y''z''$ system, the relation between the $E$-field and the modulated increment of the local displacement field $\boldsymbol{D} = D_x\hat{\boldsymbol{x}} + D_y\hat{\boldsymbol{y}} + D_z\hat{\boldsymbol{z}}$ is written as follows:

$$\begin{pmatrix} \eta_{xx} & \eta_{xy} & \eta_{xz} \\ \eta_{yx} & \eta_{yy} & \eta_{yz} \\ \eta_{zx} & \eta_{zy} & \eta_{zz} \end{pmatrix} \begin{pmatrix} D_x \\ D_y \\ D_z \end{pmatrix} = \begin{pmatrix} \varepsilon_{h1} & 0 & 0 \\ 0 & \varepsilon_{h2} & 0 \\ 0 & 0 & \varepsilon_{h3} \end{pmatrix} \begin{pmatrix} \eta_{xx} & \eta_{xy} & \eta_{xz} \\ \eta_{yx} & \eta_{yy} & \eta_{yz} \\ \eta_{zx} & \eta_{zy} & \eta_{zz} \end{pmatrix} \begin{pmatrix} E_x \\ E_y \\ E_z \end{pmatrix}. \tag{A4}$$

Considering that the $3 \times 3$ matrix $\tilde{\eta}$ is unitary, its inverse $\tilde{\eta}^{-1}$ equals its transpose $\tilde{\eta}^T$, in which case Eq.(A4) yields

$$\begin{pmatrix} D_x \\ D_y \\ D_z \end{pmatrix} = \underbrace{\overbrace{\begin{pmatrix} \eta_{xx} & \eta_{yx} & \eta_{zx} \\ \eta_{xy} & \eta_{yy} & \eta_{zy} \\ \eta_{xz} & \eta_{yz} & \eta_{zz} \end{pmatrix}}^{\tilde{\eta}^T} \begin{pmatrix} \varepsilon_{h1} & 0 & 0 \\ 0 & \varepsilon_{h2} & 0 \\ 0 & 0 & \varepsilon_{h3} \end{pmatrix} \overbrace{\begin{pmatrix} \eta_{xx} & \eta_{xy} & \eta_{xz} \\ \eta_{yx} & \eta_{yy} & \eta_{yz} \\ \eta_{zx} & \eta_{zy} & \eta_{zz} \end{pmatrix}}^{\tilde{\eta}}}_{\tilde{\varepsilon}_h} \begin{pmatrix} E_x \\ E_y \\ E_z \end{pmatrix}. \tag{A5}$$

Hereinafter, the $3 \times 3$ coefficient matrix relating the incremental $D$-field to the local $E$-field in accordance with Eq.(A5) will be denoted by the dielectric tensor $\tilde{\varepsilon}_h$, and the various elements of $\tilde{\varepsilon}_h$ will be identified as $\varepsilon_{xx}$, $\varepsilon_{xy}$, $\varepsilon_{xz}$ (1st row), $\varepsilon_{yx}$, $\varepsilon_{yy}$, $\varepsilon_{yz}$ (2nd row), and $\varepsilon_{zx}$, $\varepsilon_{zy}$, $\varepsilon_{zz}$ (3rd row).

## Appendix B
## The Fourier Transform of a Trapezoidal Grating Profile

Occasionally, when the grating profile is sufficiently simple, the Fourier coefficients $g_{mn}$ of the function $g(x,z)$, which is a periodic function of $x$ (period $= \Lambda$), can be analytically determined. This is the case, for instance, for the trapezoidal index profile shown in Fig. B1. Here the corners of the trapezoid in the $xz$-plane are located at $(x,z) = (\alpha, 0)$, $(\beta, -d)$, $(\gamma, -d)$, and $(\Lambda, 0)$. The uniformly birefringent region of the grating is marked with an arrow—which could be representative of the optical axis of a uniaxial birefringent medium. Throughout the entire slab, the host material has a uniform dielectric constant $\varepsilon_2(\omega)$, which is augmented by a dielectric tensor $\tilde{\varepsilon}_h(\omega)$ within the birefringent region. Defining the function $g(x,z)$ such that it equals 1.0 inside the birefringent region, and zero elsewhere, we may write the incremental contribution of birefringence to the overall dielectric tensor of the hologram as $\tilde{\varepsilon}_h g(x,z)$. In general, the Fourier series coefficients $g_{mn}$ of the two-dimensional function $g(x,z)$ are evaluated as follows:

$$g_{mn} = (\Lambda d)^{-1} \int_{z=-d}^{0} \int_{x=0}^{\Lambda} g(x,z) \exp[-\mathrm{i}2\pi(mx/\Lambda + nz/d)]\,\mathrm{d}x\mathrm{d}z. \qquad \text{(B1)}$$

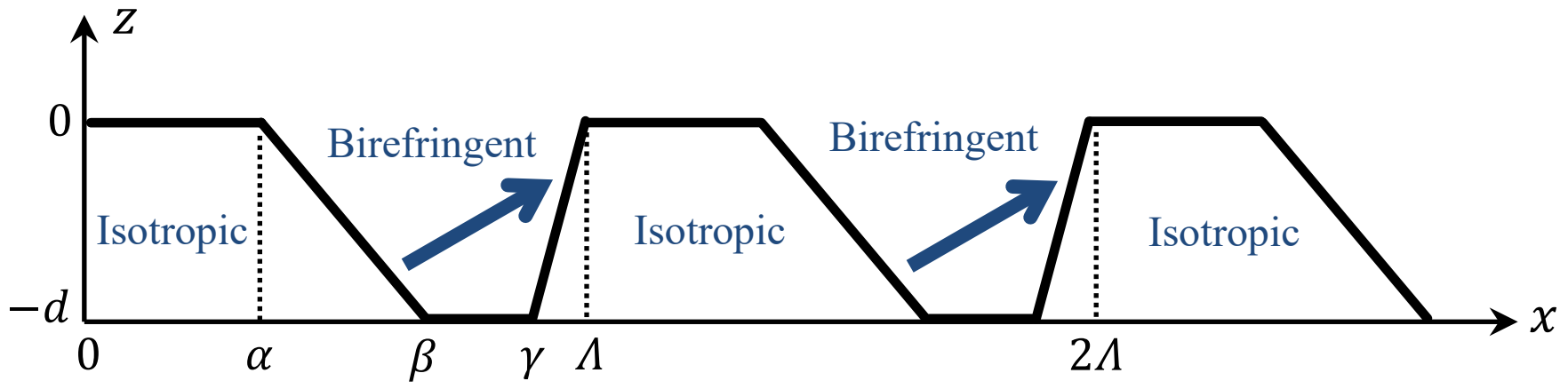

**Fig. B1**. A trapezoidal grating consisting of a birefringent region embedded within a homogeneous and isotropic host medium of dielectric constant $\varepsilon_2$. The holographic slab has thickness $d$, and the grating period along the $x$-axis is $\Lambda$. The corners of the trapezoid are located at $(x,z) = (\alpha,0)$, $(\beta,-d)$, $(\gamma,-d)$, and $(\Lambda,0)$. The function $g(x,z)$ is defined such that it equals 1.0 inside the birefringent trapezoids, and zero elsewhere.

Below, we evaluate the trapezoidal grating's Fourier coefficients $g_{0,0}$, $g_{m,0}$, $g_{0,n}$, and $g_{m,n}$ for the case of $m \neq 0$ and $n \neq 0$.

$$g_{0,0} = ½[1 + (\gamma - \alpha - \beta)/\Lambda]. \qquad \text{(B2)}$$

$$g_{m,0} = (\Lambda d)^{-1} \int_{z=-d}^{0} \int_{x=\alpha-(\beta-\alpha)z/d}^{\Lambda+(\Lambda-\gamma)z/d} \exp(-\mathrm{i}2\pi m x/\Lambda)\,\mathrm{d}x\mathrm{d}z$$

$$= \frac{\mathrm{i}}{2\pi m d} \int_{z=-d}^{0} \exp\{-\mathrm{i}2\pi m[1 + (1-\gamma/\Lambda)z/d]\} - \exp\{-\mathrm{i}2\pi m[\alpha - (\beta-\alpha)z/d]/\Lambda\}\,\mathrm{d}z$$

$$= \frac{\mathrm{i}}{2\pi m d} \times \frac{\mathrm{i}d}{2\pi m(1-\gamma/\Lambda)} \exp(-\mathrm{i}2\pi m)\{1 - \exp[\mathrm{i}2\pi m(1-\gamma/\Lambda)]\}$$

$$- \frac{\mathrm{i}}{2\pi m d} \times \frac{\Lambda d}{\mathrm{i}2\pi m(\beta-\alpha)} \exp(-\mathrm{i}2\pi m\alpha/\Lambda)\{1 - \exp[-\mathrm{i}2\pi m(\beta-\alpha)/\Lambda]\}$$

$$= \frac{\exp(-\mathrm{i}2\pi m\beta/\Lambda) - \exp(-\mathrm{i}2\pi m\alpha/\Lambda)}{(2\pi m)^2(\beta-\alpha)/\Lambda} - \frac{1 - \exp(-\mathrm{i}2\pi m\gamma/\Lambda)}{(2\pi m)^2(1-\gamma/\Lambda)}. \qquad \text{(B3)}$$

In special cases when $\alpha = \beta$ and/or $\gamma = \Lambda$, the singularities of the above expression can be removed via a Taylor series expansion of the numerator(s). For instance, factoring out the term $\exp(-\mathrm{i}2\pi m\alpha/\Lambda)$ leaves $\exp[-\mathrm{i}2\pi m(\beta-\alpha)/\Lambda] - 1$ in the numerator of the first term on the right-hand side of Eq.(B3). In the limit when $\beta \to \alpha$, this numerator approaches $-\mathrm{i}2\pi m(\beta-\alpha)/\Lambda$, which simplifies the first term on the right-hand side of the equation to $[\exp(-\mathrm{i}2\pi m\alpha/\Lambda)]/(\mathrm{i}2\pi m)$.

$$
\begin{aligned}
g_{0,n} &= (\Lambda d)^{-1}\int_{z=-d}^{0}\int_{x=\alpha-(\beta-\alpha)z/d}^{\Lambda+(\Lambda-\gamma)z/d}\exp(-\mathrm{i}2\pi nz/d)\,\mathrm{d}x\mathrm{d}z\\
&= (\Lambda d)^{-1}\int_{z=-d}^{0}[(\Lambda-\alpha)+(\Lambda-\gamma+\beta-\alpha)z/d]\exp(-\mathrm{i}2\pi nz/d)\,\mathrm{d}z\\
&= \frac{\mathrm{i}(\Lambda-\alpha)[1-\exp(\mathrm{i}2\pi n)]}{2\pi n\Lambda}+[1+(\beta-\alpha-\gamma)/\Lambda]\int_{-1}^{0}\zeta\exp(-\mathrm{i}2\pi n\zeta)\,\mathrm{d}\zeta \quad \leftarrow \text{Integration by parts} \quad (\exp(\mathrm{i}2\pi n)\to 1.0)\\
&= [1+(\beta-\alpha-\gamma)/\Lambda]\left\{\frac{\mathrm{i}\zeta\exp(-\mathrm{i}2\pi n\zeta)}{2\pi n}\bigg|_{-1}^{0}-\frac{\mathrm{i}}{2\pi n}\int_{-1}^{0}\exp(-\mathrm{i}2\pi n\zeta)\,\mathrm{d}\zeta\right\}\\
&= [1+(\beta-\alpha-\gamma)/\Lambda]\left\{\frac{\mathrm{i}\exp(\mathrm{i}2\pi n)}{2\pi n}+\frac{1-\exp(\mathrm{i}2\pi n)}{(2\pi n)^2}\right\} \quad (\exp(\mathrm{i}2\pi n)\to 1.0)\\
&= \frac{\mathrm{i}[1+(\beta-\alpha-\gamma)/\Lambda]}{2\pi n}.
\end{aligned}
\tag{B4}
$$

Finally, the general expression for $g_{mn}$ when both $m$ and $n$ are nonzero is determined as follows:

$$
\begin{aligned}
g_{mn} &= (\Lambda d)^{-1}\int_{z=-d}^{0}\int_{x=\alpha-(\beta-\alpha)z/d}^{\Lambda+(\Lambda-\gamma)z/d}\exp[-\mathrm{i}2\pi(mx/\Lambda+nz/d)]\,\mathrm{d}x\mathrm{d}z\\
&= \frac{\mathrm{i}}{2\pi md}\int_{z=-d}^{0}\exp[-\mathrm{i}2\pi m-\mathrm{i}2\pi(n+m-m\gamma/\Lambda)z/d]\\
&\qquad -\exp\{-\mathrm{i}2\pi m\alpha/\Lambda-\mathrm{i}2\pi[n-m(\beta-\alpha)/\Lambda]z/d\}\,\mathrm{d}z\\
&= \frac{\mathrm{i}}{2\pi md}\left\{\exp(-\mathrm{i}2\pi m)\times\frac{\mathrm{i}d\{1-\exp[\mathrm{i}2\pi(n+m-m\gamma/\Lambda)]\}}{2\pi(n+m-m\gamma/\Lambda)}\right.\\
&\qquad \left. -\exp(-\mathrm{i}2\pi m\alpha/\Lambda)\times\frac{\mathrm{i}d(1-\exp\{\mathrm{i}2\pi[n-m(\beta-\alpha)/\Lambda]\})}{2\pi[n-m(\beta-\alpha)/\Lambda]}\right\}\\
&= \frac{\exp(-\mathrm{i}2\pi m\alpha/\Lambda)-\exp[\mathrm{i}2\pi(n-m\beta/\Lambda)]}{(2\pi)^2m[n-m(\beta-\alpha)/\Lambda]}-\frac{\exp(-\mathrm{i}2\pi m)-\exp[\mathrm{i}2\pi(n-m\gamma/\Lambda)]}{(2\pi)^2m(n+m-m\gamma/\Lambda)}\\
&= \frac{\exp(-\mathrm{i}2\pi m\alpha/\Lambda)-\exp(-\mathrm{i}2\pi m\beta/\Lambda)}{(2\pi)^2m[n-m(\beta-\alpha)/\Lambda]}-\frac{1-\exp(-\mathrm{i}2\pi m\gamma/\Lambda)}{(2\pi)^2m(n+m-m\gamma/\Lambda)}.
\end{aligned}
\tag{B5}
$$

It is readily seen that, upon setting $n=0$, the Fourier coefficient $g_{mn}$ of Eq.(B5) reduces to $g_{m,0}$ of Eq.(B3). Derivation of the coefficient $g_{0,n}$ appearing in Eq.(B4) as the limit of Eq.(B5) when $m\to 0$ is also possible (with the aid of Taylor series expansions), but it requires further attention to detail. Specifically, in the penultimate line of Eq.(B5), we must set $\exp(\mathrm{i}2\pi n)$ to zero, then treat the parameter $m$ appearing in the exponentials as a small (but non-zero) entity, in which case $\exp(-\mathrm{i}2\pi m\alpha/\Lambda)\to 1-\mathrm{i}2\pi m\alpha/\Lambda$, with similar expressions holding for the other exponential functions as well. We will then have

$$
\begin{aligned}
\lim_{m\to 0} g_{mn} &= \lim_{m\to 0}\left\{\frac{-\mathrm{i}2\pi m\alpha/\Lambda+\mathrm{i}2\pi m\beta/\Lambda}{(2\pi)^2m[n-m(\beta-\alpha)/\Lambda]}-\frac{-\mathrm{i}2\pi m+\mathrm{i}2\pi m\gamma/\Lambda}{(2\pi)^2m(n+m-m\gamma/\Lambda)}\right\}\\
&= \lim_{m\to 0}\left\{\frac{\mathrm{i}(\beta-\alpha)/\Lambda}{2\pi[n-\cancel{m}(\beta-\alpha)/\Lambda]}+\frac{\mathrm{i}(1-\gamma/\Lambda)}{2\pi(n+\cancel{m}-\cancel{m}\gamma/\Lambda)}\right\}=\frac{\mathrm{i}[1+(\beta-\alpha-\gamma)/\Lambda]}{2\pi n}=g_{0,n}.
\end{aligned}
\tag{B6}
$$

Finally, it must be pointed out that, when $\alpha=\beta$, the first term on the right-hand side of Eq.(B5) vanishes, and when $\gamma=\Lambda$, the second term vanishes, but the overall expression for $g_{mn}$ does *not* become singular—considering that both $m$ and $n$ in Eq.(B5) are assumed to be non-zero. Such singularities arise only in conjunction with the Fourier coefficient $g_{m,0}$ of Eq.(B3), which would subsequently require proper treatment, as described earlier.

## Appendix C

### Matching the Boundary Conditions at the Upper and Lower Facets of the Hologram

**Incident beam**: Since the incident plane-wave propagates along the negative $z$-axis, we must choose $90° < \theta_0 < 180°$. We will then have

$k_{z0} = \cos\theta_0 < 0$

$$\boldsymbol{k}^{(\mathrm{inc})} = k_{x0}\hat{\boldsymbol{x}} + k_{y0}\hat{\boldsymbol{y}} + k_{z0}\hat{\boldsymbol{z}} = (2\pi n_1/\lambda_0)(\sin\theta_0\cos\varphi_0\,\hat{\boldsymbol{x}} + \sin\theta_0\sin\varphi_0\,\hat{\boldsymbol{y}} + \cos\theta_0\,\hat{\boldsymbol{z}}). \tag{C1}$$

$$\boldsymbol{E}^{(\mathrm{inc})} = (E_{x0}\hat{\boldsymbol{x}} + E_{y0}\hat{\boldsymbol{y}} + E_{z0}\hat{\boldsymbol{z}})\exp[\mathrm{i}(k_{x0}x + k_{y0}y + k_{z0}z)]. \tag{C2}$$

$$\boldsymbol{H}^{(\mathrm{inc})} = (H_{x0}\hat{\boldsymbol{x}} + H_{y0}\hat{\boldsymbol{y}} + H_{z0}\hat{\boldsymbol{z}})\exp[\mathrm{i}(k_{x0}x + k_{y0}y + k_{z0}z)]. \tag{C3}$$

Maxwell's 1st equation: $\boldsymbol{k}_0 \cdot \boldsymbol{E}_0 = 0 \quad \rightarrow \quad E_{z0} = -(k_{x0}E_{x0} + k_{y0}E_{y0})/k_{z0}.$ (C4)

Maxwell's 3rd equation: $\boldsymbol{k}_0 \times \boldsymbol{E}_0 = \mu_0\omega\boldsymbol{H}_0 \quad \rightarrow \quad \begin{pmatrix} H_{x0} \\ H_{y0} \end{pmatrix} = (\mu_0\omega)^{-1}\begin{pmatrix} k_{y0}E_{z0} - k_{z0}E_{y0} \\ k_{z0}E_{x0} - k_{x0}E_{z0} \end{pmatrix}$

$\omega = 2\pi c/\lambda_0$

$$\rightarrow \begin{pmatrix} Z_0H_{x0}^{(\mathrm{inc})} \\ Z_0H_{y0}^{(\mathrm{inc})} \end{pmatrix} = (\omega k_{z0}/c)^{-1}\begin{pmatrix} -k_{x0}k_{y0} & -k_{y0}^2 - k_{z0}^2 \\ k_{x0}^2 + k_{z0}^2 & k_{x0}k_{y0} \end{pmatrix}\begin{pmatrix} E_{x0}^{(\mathrm{inc})} \\ E_{y0}^{(\mathrm{inc})} \end{pmatrix} = \begin{pmatrix} \eta_{11}^{(\mathrm{inc})} & \eta_{12}^{(\mathrm{inc})} \\ \eta_{21}^{(\mathrm{inc})} & \eta_{22}^{(\mathrm{inc})} \end{pmatrix}\begin{pmatrix} E_{x0}^{(\mathrm{inc})} \\ E_{y0}^{(\mathrm{inc})} \end{pmatrix}. \tag{C5}$$

**Reflected beam of order $\mu$**: $k_{z\mu}^{(\mathrm{ref})} = \sqrt{(2\pi n_1/\lambda_0)^2 - (k_{x0} + \mu K_x)^2 - k_{y0}^2}$. When $k_{z\mu}$ is real, choose the positive sign for the square root, because all the various reflected beams propagate in the positive $z$-direction. When $k_{z\mu}$ is imaginary, it must be placed in the upper-half of the complex plane, since the corresponding evanescent waves must decay along the positive $z$-axis.

$$\boldsymbol{E}^{(\mathrm{ref})} = \left(E_{x\mu}\hat{\boldsymbol{x}} + E_{y\mu}\hat{\boldsymbol{y}} + E_{z\mu}\hat{\boldsymbol{z}}\right)\exp\{\mathrm{i}[(k_{x0} + \mu K_x)x + k_{y0}y + k_{z\mu}^{(\mathrm{ref})}z]\}. \tag{C6}$$

$$\boldsymbol{H}^{(\mathrm{ref})} = \left(H_{x\mu}\hat{\boldsymbol{x}} + H_{y\mu}\hat{\boldsymbol{y}} + H_{z\mu}\hat{\boldsymbol{z}}\right)\exp\{\mathrm{i}[(k_{x0} + \mu K_x)x + k_{y0}y + k_{z\mu}^{(\mathrm{ref})}z]\}. \tag{C7}$$

Once again, Eqs.(C4) and (C5) apply to all reflected plane-waves, provided that $k_{x0}$ and $k_{z0}$ are replaced with $k_{x0} + \mu K_x$ and $k_{z\mu}^{(\mathrm{ref})}$, respectively. A streamlined version of Eq.(C5) will be

$$\begin{pmatrix} Z_0H_{x\mu}^{(\mathrm{ref})} \\ Z_0H_{y\mu}^{(\mathrm{ref})} \end{pmatrix} = \begin{pmatrix} \eta_{\mu,11}^{(\mathrm{ref})} & \eta_{\mu,12}^{(\mathrm{ref})} \\ \eta_{\mu,21}^{(\mathrm{ref})} & \eta_{\mu,22}^{(\mathrm{ref})} \end{pmatrix}\begin{pmatrix} E_{x\mu}^{(\mathrm{ref})} \\ E_{y\mu}^{(\mathrm{ref})} \end{pmatrix}, \tag{C8}$$

where

$$\eta_{\mu,11}^{(\mathrm{ref})} = -(\lambda_0/2\pi)\,(k_{x0} + \mu K_x)k_{y0}/k_{z\mu}, \tag{C8a}$$

$$\eta_{\mu,12}^{(\mathrm{ref})} = -(\lambda_0/2\pi)\,(k_{y0}^2 + k_{z\mu}^2)/k_{z\mu}, \tag{C8b}$$

$$\eta_{\mu,21}^{(\mathrm{ref})} = (\lambda_0/2\pi)\left[(k_{x0} + \mu K_x)^2 + k_{z\mu}^2\right]/k_{z\mu}, \tag{C8c}$$

$$\eta_{\mu,22}^{(\mathrm{ref})} = (\lambda_0/2\pi)\,(k_{x0} + \mu K_x)k_{y0}/k_{z\mu}. \tag{C8d}$$

In the above equations, $c = 1/\sqrt{\mu_0\varepsilon_0} = 299{,}792{,}458\ \mathrm{m/sec}$, and $Z_0 = \sqrt{\mu_0/\varepsilon_0} \cong 376.730313\ \Omega$.

**Transmitted beam of order $\boldsymbol{\mu}$**: $k_{z\mu}^{(\text{trans})} = -\sqrt{(2\pi n_3/\lambda_0)^2 - (k_{x0} + \mu K_x)^2 - k_{y0}^2}$. When $k_{z\mu}$ is real, choose the negative sign for the square root, because all the various transmitted beams propagate in the negative $z$-direction. When $k_{z\mu}$ is imaginary, it must be placed in the lower-half of the complex plane, since the corresponding evanescent waves must decay along the negative $z$-axis. The $\boldsymbol{E}$ and $\boldsymbol{H}$ fields of the transmitted plane-waves are similar to those given by Eqs.(C6) and (C7), provided that $k_{z\mu}^{(\text{trans})}$ is substituted for $k_{z\mu}^{(\text{ref})}$. For the transmitted plane-waves, we will assume that the origin of coordinates is at $z = 0$ (rather than at $z = -d$). This simplifies the following equations without changing the physical situation. In effect, we are absorbing an extra factor of $\exp[-\mathrm{i}k_{z\mu}^{(\text{trans})}d]$ into the transmitted field amplitudes $E_{x\mu}$, $E_{y\mu}$, $H_{x\mu}$, and $H_{y\mu}$.

**Matching the boundary conditions**: At the upper interface, $z = 0$, we have a single incident beam with known $k_{x0}, k_{y0}, k_{z0}, E_{x0}, E_{y0}$, as well as $H_{x0}, H_{y0}$ given by Eq.(C5). We also have $\Gamma = \Bbbk_1\Bbbk_2$ reflected plane-waves (i.e., $\Bbbk_1$ distinct plane-waves, each with a degeneracy of $\Bbbk_2$), each plane-wave having two unknown coefficients, $E_{x\mu}^{(\text{ref})}$ and $E_{y\mu}^{(\text{ref})}$. The tangential $H$-field components, namely, $H_{x\mu}^{(\text{ref})}$ and $H_{y\mu}^{(\text{ref})}$, are related to the corresponding $E$-field components via Eqs.(C8). At this boundary, we must write $4\Gamma$ continuity equations for all the tangential $E$- and $H$-field components.

A similar situation arises at the lower interface at $z = -d$, where there exist $\Gamma = \Bbbk_1\Bbbk_2$ transmitted plane-waves (again, $\Bbbk_1$ distinct plane-waves, each with a degeneracy of $\Bbbk_2$), each plane-wave having two unknown coefficients, $E_{x\mu}^{(\text{trans})}$ and $E_{y\mu}^{(\text{trans})}$. Here, the tangential $H$-field amplitudes, $H_{x\mu}^{(\text{trans})}$ and $H_{y\mu}^{(\text{trans})}$, are related to the corresponding $E$-field amplitudes via Eqs.(C8), albeit with $k_{z\mu}^{(\text{trans})}$ substituting for $k_{z\mu}^{(\text{ref})}$. Again, we must write $4\Gamma$ continuity equations at the lower $(z = -d)$ interface.

Inside the hologram, $-d \le z \le 0$, there reside $4\Gamma^2$ plane-waves, corresponding to the $4\Gamma$ eigenvalues and eigenvectors $(k_{zm}, V_m)$ obtained from Eq.(14). Each eigenvector $V_m$, having $4\Gamma$ elements (i.e., $E_x, E_y, Z_0H_x, Z_0H_y$ for each one of $\Gamma$ plane-waves) and written in expanded form as $(v_{m,1}, v_{m,2}, \cdots, v_{m,4\Gamma})^T$, may be multiplied by an arbitrary scale-factor $\zeta_m$ and continue to be a solution of our coupled-wave equations. We thus have $4\Gamma$ additional unknown coefficients $\zeta_1, \zeta_2, \zeta_3, \cdots, \zeta_{4\Gamma}$. Inside the hologram, all the field amplitudes $E_{x\mu}, E_{y\mu}, Z_0H_{x\mu}, Z_0H_{y\mu}$ at the upper interface $(z = 0)$ are now given by $\zeta_m V_m$, whereas, at the bottom interface $(z = -d)$, these amplitudes must be multiplied by the complex coefficient $\tau_{m,\nu} = \exp[-\mathrm{i}(k_{zm} + \nu K_z)d]$. Given that $\nu K_z d = 2\pi\nu$ is an integer-multiple of $2\pi$, the coefficients $\tau_{m,\nu}$ become independent of $\nu$, thus allowing us to write them simply as $\tau_m = \exp(-\mathrm{i}k_{zm}d)$.

All in all, we have $8\Gamma$ equations in $8\Gamma$ unknowns. The unknowns can be arranged in an $8\Gamma \times 1$ column vector, as follows:

$$\Big[\overbrace{\cdots, E_{x\mu}^{(\text{ref})}, E_{y\mu}^{(\text{ref})}, \cdots}^{2\Gamma}, \overbrace{\cdots, E_{x\mu}^{(\text{trans})}, E_{y\mu}^{(\text{trans})}, \cdots}^{2\Gamma}, \overbrace{\zeta_1, \zeta_2, \zeta_3, \cdots, \zeta_{4\Gamma}}^{4\Gamma}\Big]^T. \tag{C9}$$

The $8\Gamma$ linear equations obtained by matching the tangential $E$- and $H$-field components at the two boundaries can now be written as the $8\Gamma \times 8\Gamma$ matrix equation depicted in Fig.C1. The $8\Gamma \times 1$ vector on the right-hand side of Fig.C1 has only four non-zero elements, which correspond to the known entities $E_{x0}^{(\text{inc})}$, $E_{y0}^{(\text{inc})}$, $Z_0H_{x0}^{(\text{inc})}$, and $Z_0H_{y0}^{(\text{inc})}$. Whereas $E_{x0}^{(\text{inc})}, E_{y0}^{(\text{inc})}$ are arbitrary parameters that must be specified in advance, the corresponding magnetic field components $Z_0H_{x0}^{(\text{inc})}, Z_0H_{y0}^{(\text{inc})}$ should be derived from Eq.(C5).

Finally, the unknown coefficients appearing in Eq.(C9) can be determined by inverting the large $(8\Gamma \times 8\Gamma)$ matrix of Fig.C1, followed by multiplication into the sparse $8\Gamma \times 1$ column vector on the right-hand side of the figure. The solution thus obtained contains the reflected and transmitted field amplitudes $E_{x\mu}^{(\text{ref})}, E_{y\mu}^{(\text{ref})}, E_{x\mu}^{(\text{trans})}, E_{y\mu}^{(\text{trans})}$ for the various diffracted orders, as well as the eigenvector scale-factors $\zeta_1, \zeta_2, \cdots, \zeta_{4\Gamma}$ for the plane-waves that reside inside the hologram.

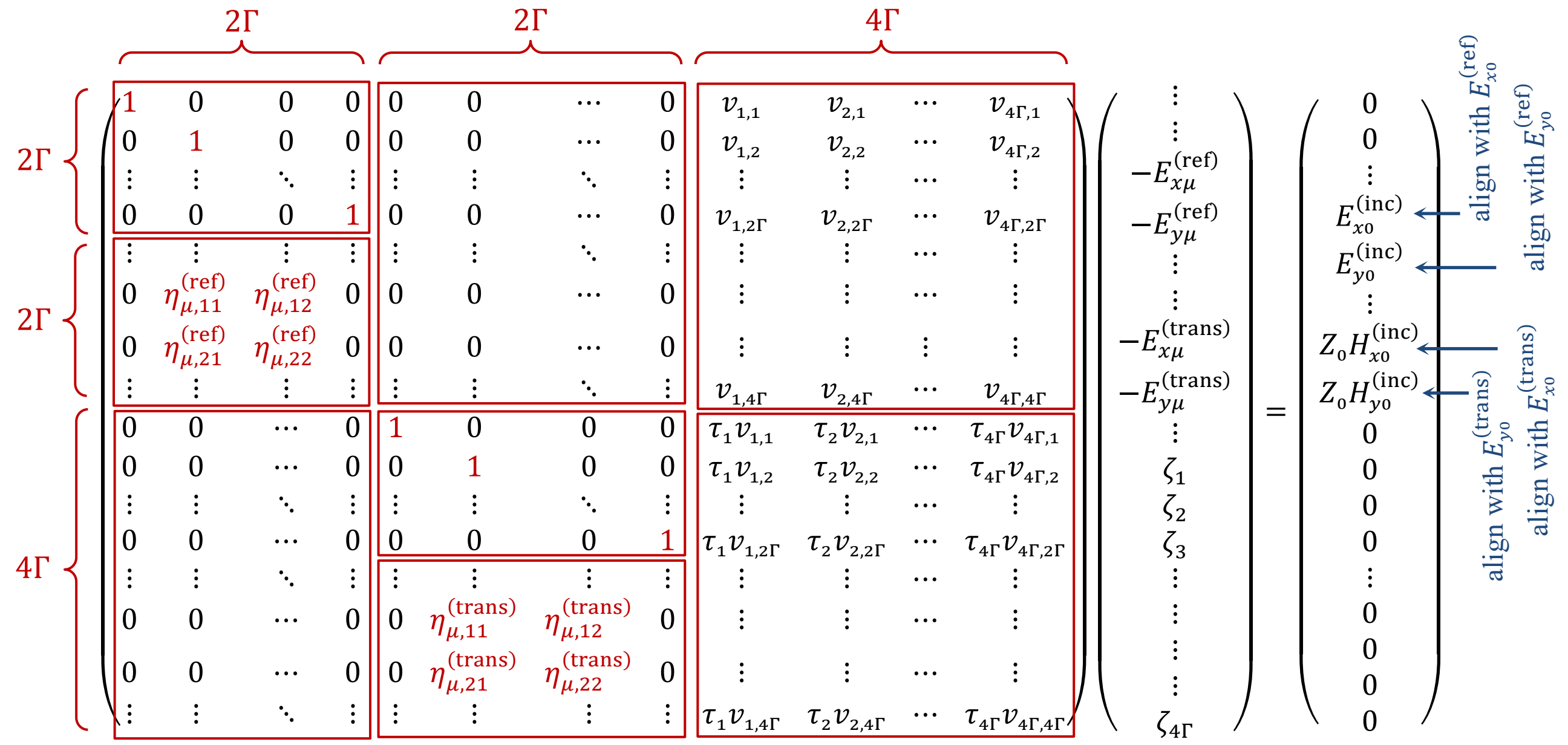

**Fig. C1**. Arrangement of $8\Gamma$ linear equations in $8\Gamma$ unknowns, obtained by matching the tangential $E$- and $H$-fields (i.e., $E_{x\mu}, E_{y\mu}, H_{x\mu}, H_{y\mu}$) at the upper and lower facets of the hologram. The $2 \times 2$ sub-matrices $[\eta_{ij}]$ are given for the reflected beams by Eqs.(C8), and for the transmitted beams by a variant of Eqs.(C8), where $k_{z\mu}^{(\text{trans})}$ must be substituted for $k_{z\mu}^{(\text{ref})}$. The $4\Gamma \times 4\Gamma$ sub-matrices on the right-hand side of the large matrix are obtained by a sequential arrangement of the eigenvectors $V_m$ of Eq.(14). The coefficients $\tau_m$ of these eigenvectors in the lower-right-hand sub-matrix are given by the complex exponential $\exp(-\mathrm{i}k_{zm}d)$, where $k_{zm}$ is the eigenvalue associated with the eigenvector $V_m = (v_{m,1}, v_{m,2}, \cdots, v_{m,4\Gamma})^T$. Two of the sub-matrices are $2\Gamma \times 2\Gamma$ identity matrices, and the two $4\Gamma \times 2\Gamma$ sub-matrices consist entirely of zeros. The $8\Gamma \times 1$ vector on the right-hand side of the equation has only four non-zero elements, which correspond to the known entities $E_{xo}^{(\text{inc})}, E_{yo}^{(\text{inc})}, Z_0 H_{xo}^{(\text{inc})}, Z_0 H_{yo}^{(\text{inc})}$. Whereas $E_{xo}^{(\text{inc})}, E_{yo}^{(\text{inc})}$ are arbitrary parameters that must be specified in advance, the corresponding magnetic field components $H_{xo}^{(\text{inc})}, H_{yo}^{(\text{inc})}$ are derived from Eq.(C5).

**Note on the method of implementation of the algorithm**. A critical problem associated with the large $(4\Gamma \times 4\Gamma)$ matrix of Eq.(14) is that, in many instances, a few eigenvalues $k_{zm}$ with *large* imaginary part appear in the computations pertaining to this equation. These eigenvalues are realisitc solutions of Maxwell's equations and must be accepted as physical entities. Proper numerical treatment of these large eigenvalues requires that we absorb the big numbers, namely, $\exp[\mathrm{Im}(k_{zm})d]$, into the corresponding $\zeta_m$ coefficients appearing in Fig.C1. In fact, for the sake of consistency, this strategy of "absorbing into the $\zeta_m$ coefficient" is implemented whenever $\mathrm{Im}(k_{zm})$ happens to be *positive*, irrespective of whether it is large or small — just so long as $\mathrm{Im}(k_{zm})$ is positive. This means that the coefficient $\tau_m$ appearing in the corresponding column $m$ of the lower-right-hand sub-matrix in Fig.C1 would become $\exp[-\mathrm{i}\mathrm{Re}(k_{zm})d]$ whenever $\mathrm{Im}(k_{zm})$ is positive (i.e., the imaginary part of $k_{zm}$ is removed from $\tau_m$). However, since $\zeta_m$ has now *absorbed* this large exponential factor, we must compensate for it by multiplying $\exp[-\mathrm{Im}(k_{zm})d]$ into the entire $m^{\text{th}}$ column of the sub-matrix appearing in the upper-right-hand corner of the big matrix in Fig.C1.

Given that $\exp[-\mathrm{Im}(k_{zm})d] < 1.0$, it will not cause any trouble with big numbers. In those instances where $\mathrm{Im}(k_{zm}) \leq 0$, no action needs to be taken, and column $m$ of the big matrix of Fig.C1 (within both sub-matrices in the upper- as well as lower-right-hand-side) remains intact. In the end, the strategy of "absorbing big numbers into the $\zeta_m$ coefficient" will solve numerical overflow problems associated with the large eigenvalues of Eq.(14) — without making any unjustified assumptions or undesirable approximations.

## Appendix D

### Conservation of Energy Among the Incident, Reflected and Transmitted Plane-Waves

In the absence of absorption inside the hologram (i.e., when both $\varepsilon_2$ and $\tilde{\varepsilon}_h$ are real-valued), conservation of energy demands that the $z$-component of the Poynting vector be preserved among the incident, reflected, and transmitted plane-waves. For monochromatic light, the time-averaged Poynting vector is given by $\langle \boldsymbol{S} \rangle = ½\mathrm{Re}(\boldsymbol{E} \times \boldsymbol{H}^*)$. Thus, for the incident plane-wave,

$$\langle S_z^{(\mathrm{inc})} \rangle = ½\mathrm{Re}\big[E_{x0}^{(\mathrm{inc})} H_{y0}^{*(\mathrm{inc})} - E_{y0}^{(\mathrm{inc})} H_{x0}^{*(\mathrm{inc})}\big]. \quad \text{(D1)}$$

In the above equation, the $E$-field components $E_{x0}^{(\mathrm{inc})}$ and $E_{y0}^{(\mathrm{inc})}$ are chosen arbitrarily at the outset, whereas the corresponding $H$-field components, $H_{x0}^{(\mathrm{inc})}$ and $H_{y0}^{(\mathrm{inc})}$, must be obtained from Eq.(C5). Note that it is the conjugated $H$-field components that appear in Eq.(D1), and that, in our chosen coordinate system of Fig.14, $\langle S_z^{(\mathrm{inc})} \rangle$ is a negative entity — i.e., the incident optical energy propagates downward.

The reflected beam consists of $N$ plane-waves corresponding to the various diffraction orders associated with the values of $\mu$. Adding up the $z$-component of the Poynting vectors of *all* the reflected plane-waves, we find the total reflected energy flux as follows:

$$\langle S_z^{(\mathrm{ref})} \rangle = \sum_{\mu} ½\mathrm{Re}\big[E_{x\mu}^{(\mathrm{ref})} H_{y\mu}^{*(\mathrm{ref})} - E_{y\mu}^{(\mathrm{ref})} H_{x\mu}^{*(\mathrm{ref})}\big]. \quad \text{(D2)}$$

The $E$-field components $E_{x\mu}^{(\mathrm{ref})}$ and $E_{y\mu}^{(\mathrm{ref})}$ in Eq.(D2) must be obtained by solving the large matrix equation shown in Fig.C1; the corresponding $H$-field components, $H_{x\mu}^{(\mathrm{ref})}$ and $H_{y\mu}^{(\mathrm{ref})}$, are subsequently derived from Eq.(C8). Note, once again, the appearance of the conjugated $H$-field components in Eq.(D2). In our chosen coordinate system, $\langle S_z^{(\mathrm{ref})} \rangle$ will turn out to be a positive entity, as the reflected optical energy of each plane-wave propagates upward. We mention in passing that, although Eq.(D2) sums over *all* reflected plane-waves, in reality, the *evanescent* waves make no contribution whatsoever to the total reflected energy flux. This is because evanescent plane-waves have an inherent 90° relative phase between $E_{x\mu}$ and $H_{y\mu}$ (also between $E_{y\mu}$ and $H_{x\mu}$ of the same diffraction order), which causes the real part of the bracketed entity on the right-hand side of Eq.(D2) to vanish.

The transmitted beam is similar to the reflected beam in many respects. It also consists of $N$ diffracted orders associated with the allowed values of $\mu$. Adding up the values of $\langle S_z^{(\mathrm{trans})} \rangle$ for all the transmitted plane-waves, we find the total transmitted energy flux, as follows:

$$\langle S_z^{(\mathrm{trans})} \rangle = \sum_{\mu} ½\mathrm{Re}\big[E_{x\mu}^{(\mathrm{trans})} H_{y\mu}^{*(\mathrm{trans})} - E_{y\mu}^{(\mathrm{trans})} H_{x\mu}^{*(\mathrm{trans})}\big]. \quad \text{(D3)}$$

The $E$-field components $E_{x\mu}^{(\text{trans})}$ and $E_{y\mu}^{(\text{trans})}$ in Eq.(D3) must be obtained by solving the large matrix equation shown in Fig.C1; the corresponding $H$-field components, $H_{x\mu}^{(\text{trans})}$ and $H_{y\mu}^{(\text{trans})}$, are subsequently derived from Eq.(C8), albeit with $k_{z\mu}^{(\text{trans})}$ substituted for $k_{z\mu}^{(\text{ref})}$. As before, the $H$-field components appearing in Eq.(D3) are conjugated, the final value of $\langle S_z^{(\text{trans})}\rangle$ turns out to be negative (as the transmitted optical energy of each plane-wave propagates downward), and the evanescent waves make no contribution to the total sum in Eq.(D3).